\documentclass[preprint,11pt]{elsarticle}

 \usepackage{epsfig}

\usepackage{amssymb}
\usepackage{graphics,amssymb,amsmath}
\usepackage{graphicx}
\usepackage{subfigure}
\newproof{pf}{Proof}
\newdefinition{rmk}{Remark}
\newtheorem{thm}{Theorem}
\newdefinition{prop}{Proposition}
\biboptions{sort&compress}
\begin{document}

\begin{frontmatter}

%% Title, authors and addresses

%% use the tnoteref command within \title for footnotes;
%% use the tnotetext command for theassociated footnote;
%% use the fnref command within \author or \address for footnotes;
%% use the fntext command for theassociated footnote;
%% use the corref command within \author for corresponding author footnotes;
%% use the cortext command for theassociated footnote;
%% use the ead command for the email address,
%% and the form \ead[url] for the home page:
%% \title{Title\tnoteref{label1}}
%% \tnotetext[label1]{}
%% \author{Name\corref{cor1}\fnref{label2}}
%% \ead{email address}
%% \ead[url]{home page}
%% \fntext[label2]{}
%% \cortext[cor1]{}
%% \address{Address\fnref{label3}}
%% \fntext[label3]{}

\title{Exponentiated Weibull-logarithmic Distribution: Model, Properties and Applications}

%% use optional labels to link authors explicitly to addresses:
%% \author[label1,label2]{}
%% \address[label1]{}
%% \address[label2]{}

\author{Eisa Mahmoudi\corref{cor1}}
\ead{emahmoudi@yazduni.ac.ir}
\author{Afsaneh Sepahdar}
\author{Artur Lemonte}
%\ead[url]{http://www.elsevier.com}

\cortext[cor1]{Corresponding author}

\address{Department of Statistics, Yazd University,
P.O. Box 89175-741, Yazd, Iran}
%\address[rvt]{Department of Statistics, Yazd University,
 %P.O. Box 89175-741, Yazd, Iran}

\begin{abstract}
In this paper, we introduce a new four-parameter generalization of
the exponentiated Weibull (EW) distribution, called the
exponentiated Weibull-logarithmic (EWL) distribution, which obtained
by compounding EW and logarithmic distributions. The new
distribution arises on a latent complementary risks scenario, in
which the lifetime associated with a particular risk is not
observable; rather, we observe only the maximum lifetime value among
all risks. The distribution exhibits decreasing, increasing,
unimodal and bathtub-shaped hazard rate functions, depending on its
parameters and contains several lifetime sub-models such as:
generalized exponential-logarithmic (GEL), complementary
Weibull-logarithmic (CWL), complementary exponential-logarithmic
(CEL), exponentiated Rayleigh-logarithmic (ERL) and
Rayleigh-logarithmic (RL) distributions. We study various properties
of the new distribution and provide numerical examples to show the
flexibility and potentiality of the model.

\end{abstract}

\begin{keyword}
EM-algorithm\sep Exponentiated Weibull distribution\sep Maximum
likelihood estimation\sep Logarithmic distribution\sep Probability
weighted moments\sep Residual life function.
%% keywords here, in the form: keyword \sep keyword

%% PACS codes here, in the form: \PACS code \sep code

 \MSC 60E05 \sep 62F10 \sep 62P99
%% or \MSC[2008] code \sep code (2000 is the default)

\end{keyword}

\end{frontmatter}

%% \linenumbers

%% main text
\section{Introduction}

%CCCCCCCCCCCCCCCCCCCCCCCCCCCCCCCCCCCCCCCCCCCCCCCCCCCCCCCCCCCCCCCCCCCCCCCCCCCCCCCCCCCCCC
%CCCCCCCCCCCCCCCCCCCCCCCCCCCCCCCCCCCCCCCCCCCCCCCCCCCCCCCCCCCCCCCCCCCCCCCCCCCCCCCCCCCCCCCC
In biological and engineering sciences study of length of organisms,
devices and materials is of major important. A substantial part of
such study is devoted to modeling the lifetime data by a failure
distribution. The Weibull and EW distributions are the most commonly
used distributions in reliability and life testing. These
distributions have several desirable properties and nice physical
interpretations. Unfortunately, however, these distributions do not
provide a reasonable parametric fit for some practical applications
where the underlying hazard functions may be decreasing, unimodal
and bathtub-shaped.

Recently, there has been a great interest among statisticians and
applied researchers in constructing flexible families of
distributions to facilitate better modeling of data. The
exponential-geometric (EG), exponential-Poisson (EP),
exponential-logarithmic (EL), exponential-power series (EPS),
Weibull-geometric (WG), Weibull-power series (WPS), exponentiated
exponential-Poisson (EEP), complementary exponential-geometric
(CEG), Poisson-exponential (PE), generalized exponential-power
series (GEPS), exponentiated Weibull-geometric (EWG) and
exponentiated Weibull-Poisson (EWP) distributions were introduced
and studied by Adamidis and Loukas \cite{Adamidis }, Kus \cite{Kus
}, Tahmasbi and Rezaei \cite{Tahmasbi }, Chahkandi and Ganjali
\cite{Chahkandi}, Barreto-Souza et al. \cite{Barreto-Souza2011 } and
Morais and Barreto-Souza \cite{Morais }, Barreto-Souza and
Cribari-Neto \cite{Barreto-Souza2009 }, Louzada-Neto et al.
\cite{Louzada}, Cancho et al. \cite{Cancho }, Mahmoudi and Jafari
\cite{Mahmoudi2011a }, Mahmoudi and Shiran \cite{Mahmoudi2011b} and
Mahmoudi and Sepahdar \cite{Mahmoudi2011c}, respectively.

In this paper, we propose a new four-parameters distribution,
referred to as the EWL distribution, which contains as special
sub-models the generalized exponential-logarithmic (GEL),
complementary Weibull-logarithmic (CWL), complementary
exponential-logarithmic (CEL), exponentiated Rayleigh-logarithmic
(ERL) and Rayleigh-logarithmic (RL) distributions, among others.
%CCCCCCCCCCCCCCCCCCCCCCCCCCCCCCCCCCCCCCCCCCCCCCCCCCCCCCCCCCCCCCCCCCCCCCCCCCCCCCCCC
%The
%main reasons for introducing the EWL distribution are: (i) This
%distribution due to its flexibility in accommodating different forms
%of the risk function is an important model that can be used in a
%variety of problems in modeling lifetime data. (ii) This
%distribution is a suitable model in a complementary risk problem
%base (Basu and Klein, \cite{Basu }) in presence of latent risks, in
%the sense that there is no information about which factor was
%responsible for the component failure and only the maximum lifetime
%value among all risks is observed. (iii) It provides a reasonable
%parametric fit to skewed data that cannot be properly fitted by
%other distributions and is a suitable model in several areas such as
%public health, actuarial science, biomedical studies, demography and
%industrial reliability.
%CCCCCCCCCCCCCCCCCCCCCCCCCCCCCCCCCCCCCCCCCCCCCCCCCCCCCCCCCCCCCCCCCCCCCCCCCCCCCCCCC
The paper is organized as follows: In Section 2, a new lifetime
distribution, called the exponentiated Weibull-logarithmic (EWL)
distribution,  is obtained by mixing exponentiated Weibull and
logarithmic distributions. Various properties of the proposed
distribution are discussed in Section 3. Estimation of the
parameters by maximum likelihood via a EM-algorithm and inference
for large sample are presented in Section 4. In Section 5, we
studied some special sub-models of the EWL distribution. Finally, in
Section 6, experimental results of the proposed distribution, based
on two real data sets, are illustrated.

%In Section 2, we review the EW distribution and its properties. In
%Section 3, we define the EWL distribution. The density, survival and
%hazard rate functions and some of their properties are given in this
%section. We derive quantiles and moments of the EWL distribution in
%Section 4. R\'{e}nyi and Shannon entropies of the EWL distribution
%are given in Section 5. Section 6 provides the moments of order
%statistics of the EWL distribution. Residual and reverse residual
%life functions of the EWL distribution are discussed in Section 7.
%In Section 8 we explain probability weighted moments. Mean
%deviations from the mean and median are obtained in Section 9.
%Section 10 is devoted to the Bonferroni and Lorenz curves of the EWL
%distribution. Estimation of the parameters by maximum likelihood via
%a EM-algorithm and inference for large sample are presented in
%Section 11. In Section 12, we studied some special sub-models of the
%EWL distribution. Applications to real data sets are given in
%Section 13 and conclusions are provided in Section 14.

%CCCCCCCCCCCCCCCCCCCCCCCCCCCCCCCCCCCCCCCCCCCCCCCCCCCCCCCCCCCCCCCCCCCCCCCCCCCCCCCCC
%CCCCCCCCCCCCCCCCCCCCCCCCCCCCCCCCCCCCCCCCCCCCCCCCCCCCCCCCCCCCCCCCCCCCCCCCCCCCCCCCCC
%CCCCCCCCCCCCCCCCCCCCCCCCCCCCCCCCCCCCCCCCCCCCCCCCCCCCCCCCCCCCCCCCCCCCCCCCCCCCCCCCCCCCCCC
\section{The proposed distribution}
Suppose that the random variable $X$ has the EW distribution where
its cdf and pdf are given by
\begin{equation}\label{cdf ExW}
F_{X}(x)=\left(1-e^{-(\beta x)^{\gamma}}\right)^{\alpha},
\end{equation}
and
\begin{equation}\label{pdf ExW}
f_X(x)=\alpha\gamma\beta^{\gamma}x^{\gamma-1}e^{-(\beta
x)^{\gamma}}\left(1-e^{-(\beta x)^{\gamma}}\right)^{\alpha-1},
\end{equation}
respectively, where $x>0$, $\alpha>0$, $\beta>0$ and $\gamma>0$.
Given $N$, let $X_{1},\cdots,X_{N}$ be independent and identify
distributed random variables from EW distribution. Let $N$ is
distributed according to the logarithmic distribution with pdf
\begin{equation*}
P(N=n)=\frac{\theta^n‎}{-n\log(1-‎\theta‎)},~n=1,2
,\cdots,~\theta>0.
\end{equation*}
Let $Y=\max(X_{1},\cdots,X_{N})$, then the conditional cdf of
$Y|N=n$ is given  by
\begin{equation}\label{dist y given N}
F_{Y|N=n}(y)=[1 - e^{-(‎\beta y‎)^{‎\gamma‎}}]^{n
‎\alpha‎},
\end{equation}
which is a EW distribution with parameters $n\alpha$, $\beta$,
$\gamma$. The EWL distribution that is defined by the marginal cdf
of $Y$, is given by
\begin{equation}\label{cdf EWL}
F(y)= ‎‎‎\frac{\log \big[1-‎\theta
‎‎‎‎‎\big(1-e^{-(‎\beta y‎)^{‎\gamma‎}} \big)
^{‎\alpha‎}  \big]}{\log \big( 1-‎\theta‎   \big)}.
\end{equation}
The pdf of EWL, denoted by EWL$(\alpha,\beta,\gamma,\theta)$, is
given by
\begin{equation}\label{pdf EWL}
f(y)=\frac{\alpha\theta\gamma\beta^{\gamma} y^{‎\gamma -1‎}
e^{-‎(\beta y‎)^{‎\gamma‎}}  \big(1-e^{-‎(\beta
y‎)^{‎\gamma‎}} \big)^{‎\alpha -1‎} ‎}{\log
(1-‎\theta‎)\big( ‎\theta‎ (1-e^{-‎(\beta
y‎)^{‎\gamma‎}} ) ^{‎\alpha‎}-1\big ) }.
\end{equation}
The graphs of EWL probability density function are displayed in Fig.
1 for selected parameter values.
\begin{figure}[t]
\centering
\includegraphics[scale=0.30]{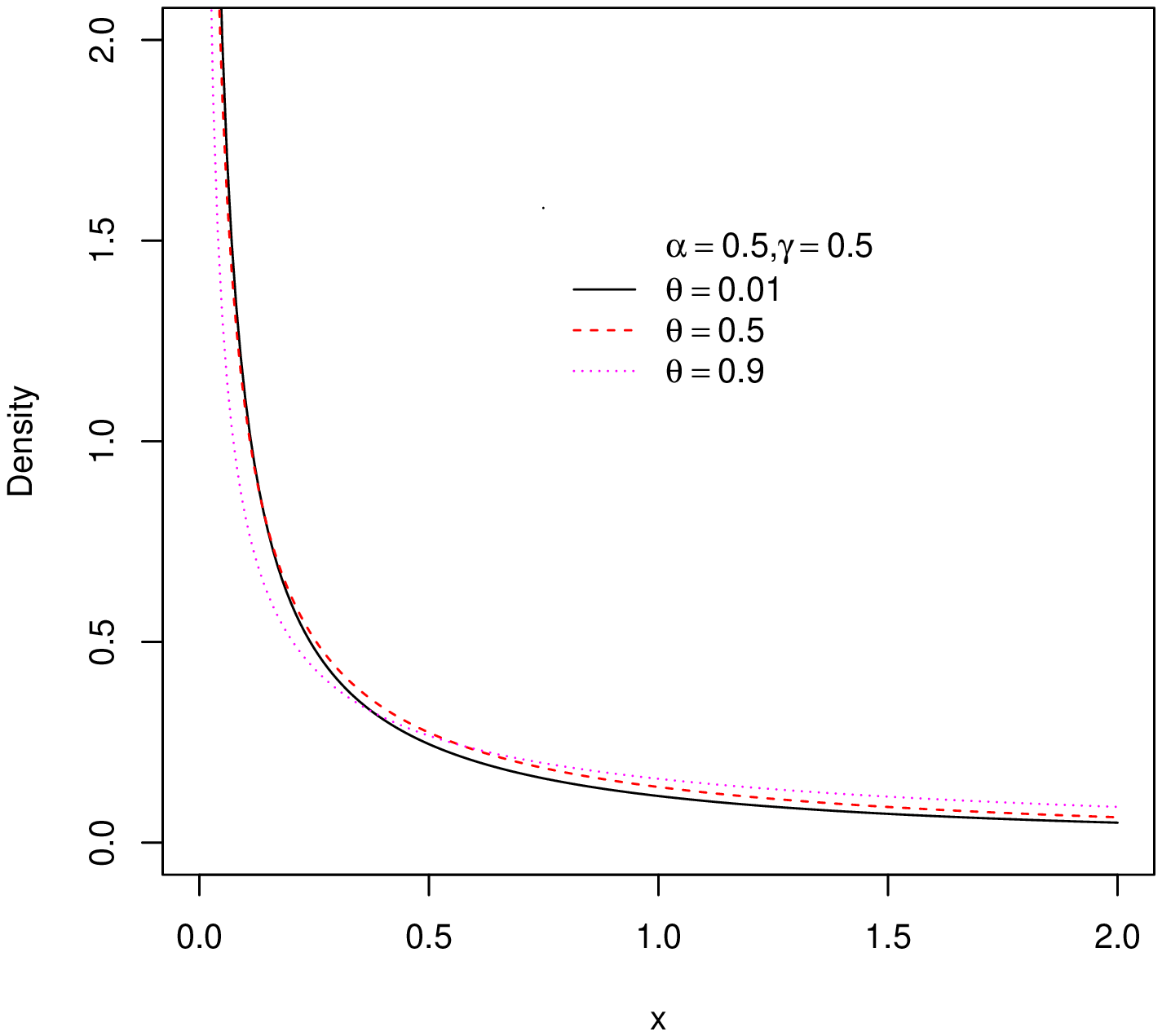}
\includegraphics[scale=0.30]{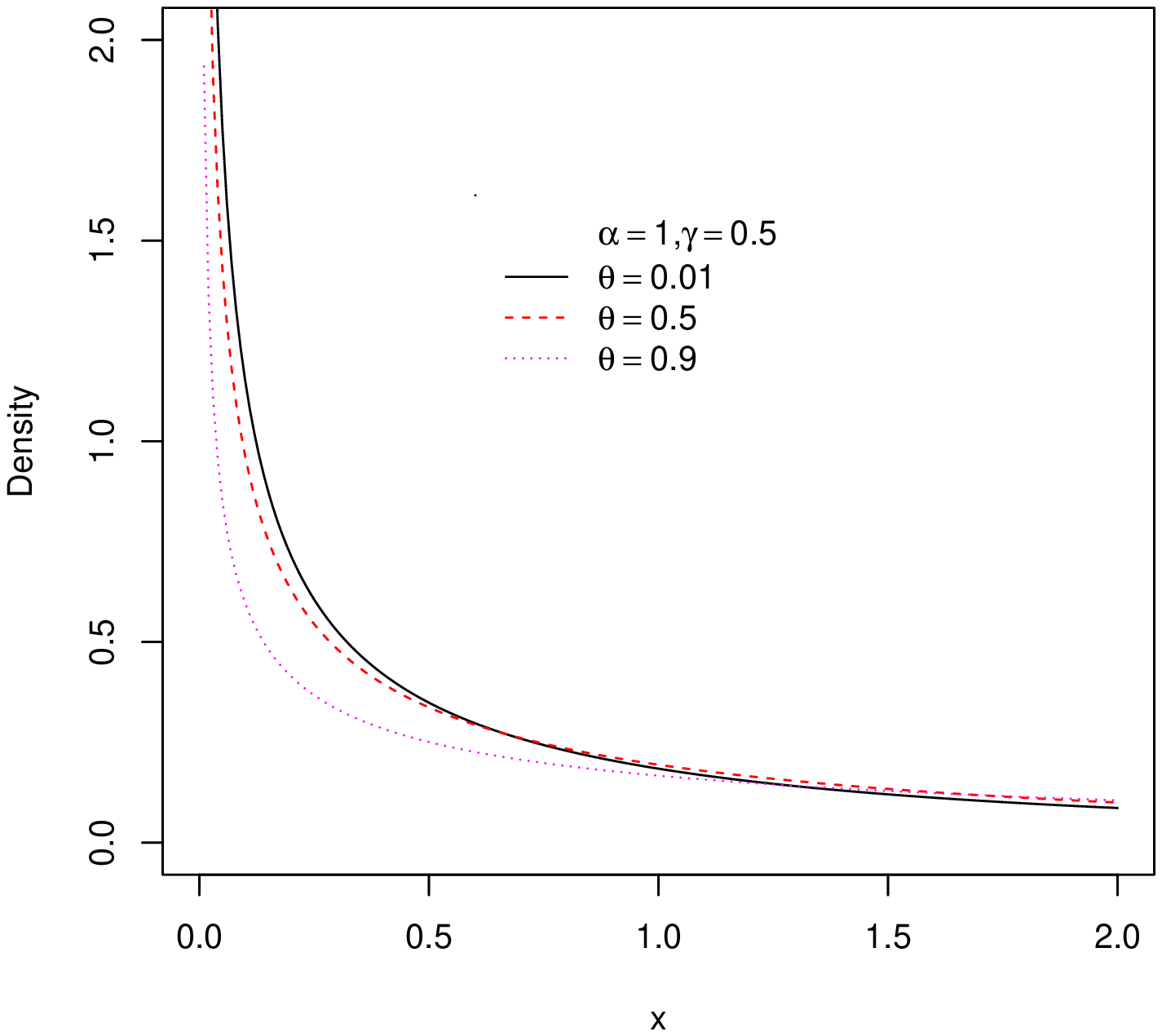}
\includegraphics[scale=0.30]{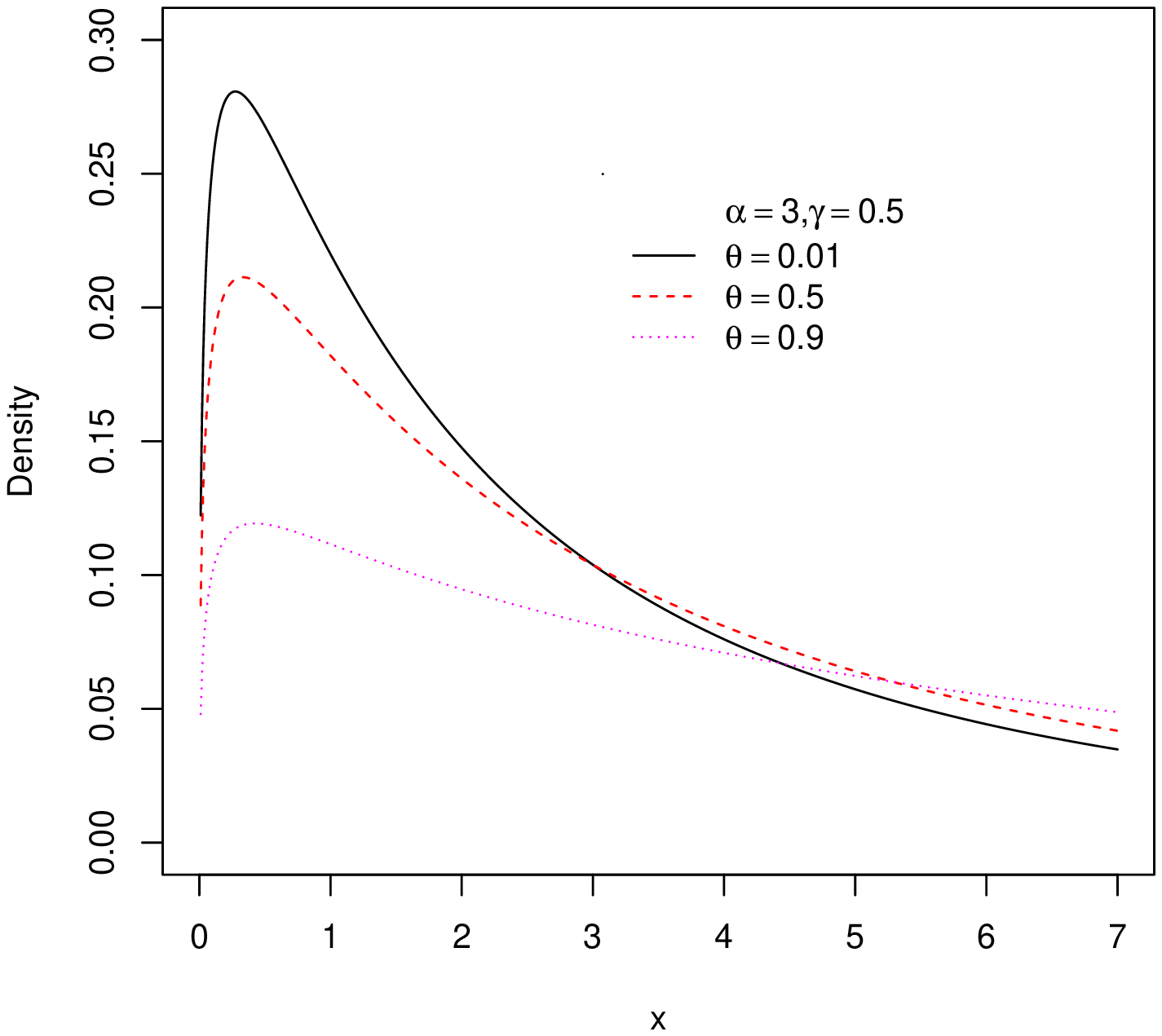}
\includegraphics[scale=0.30]{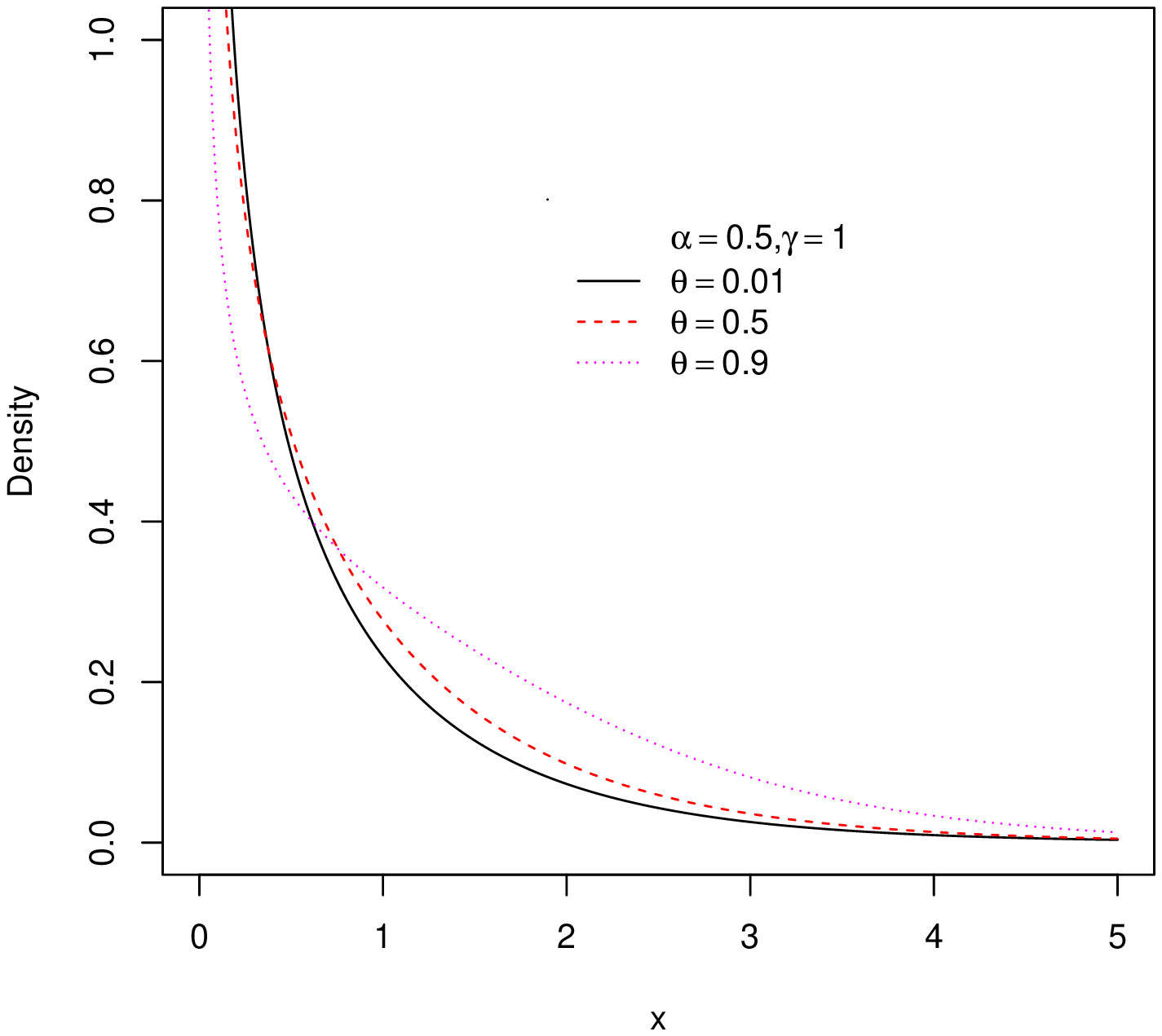}
\includegraphics[scale=0.30]{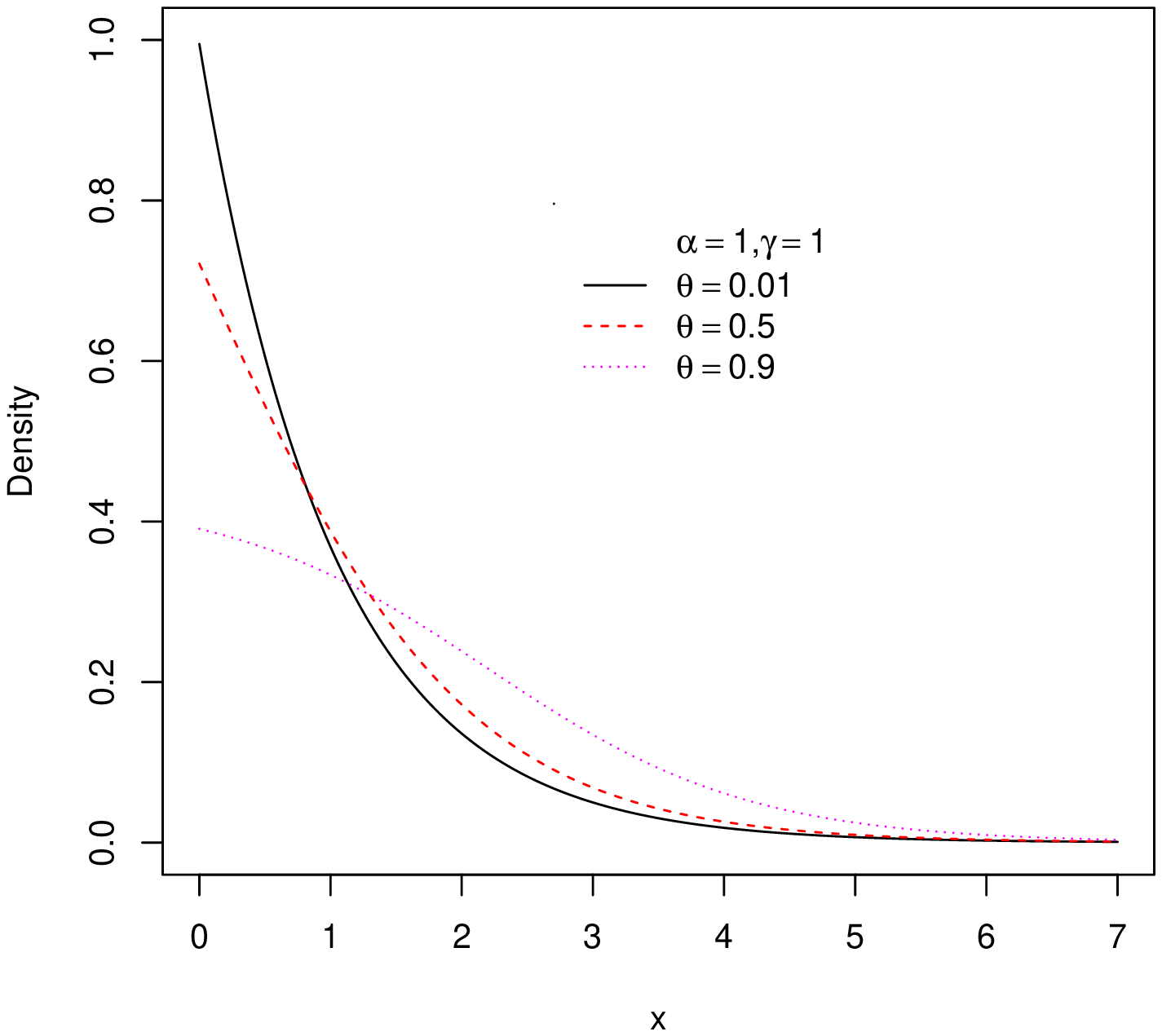}
\includegraphics[scale=0.30]{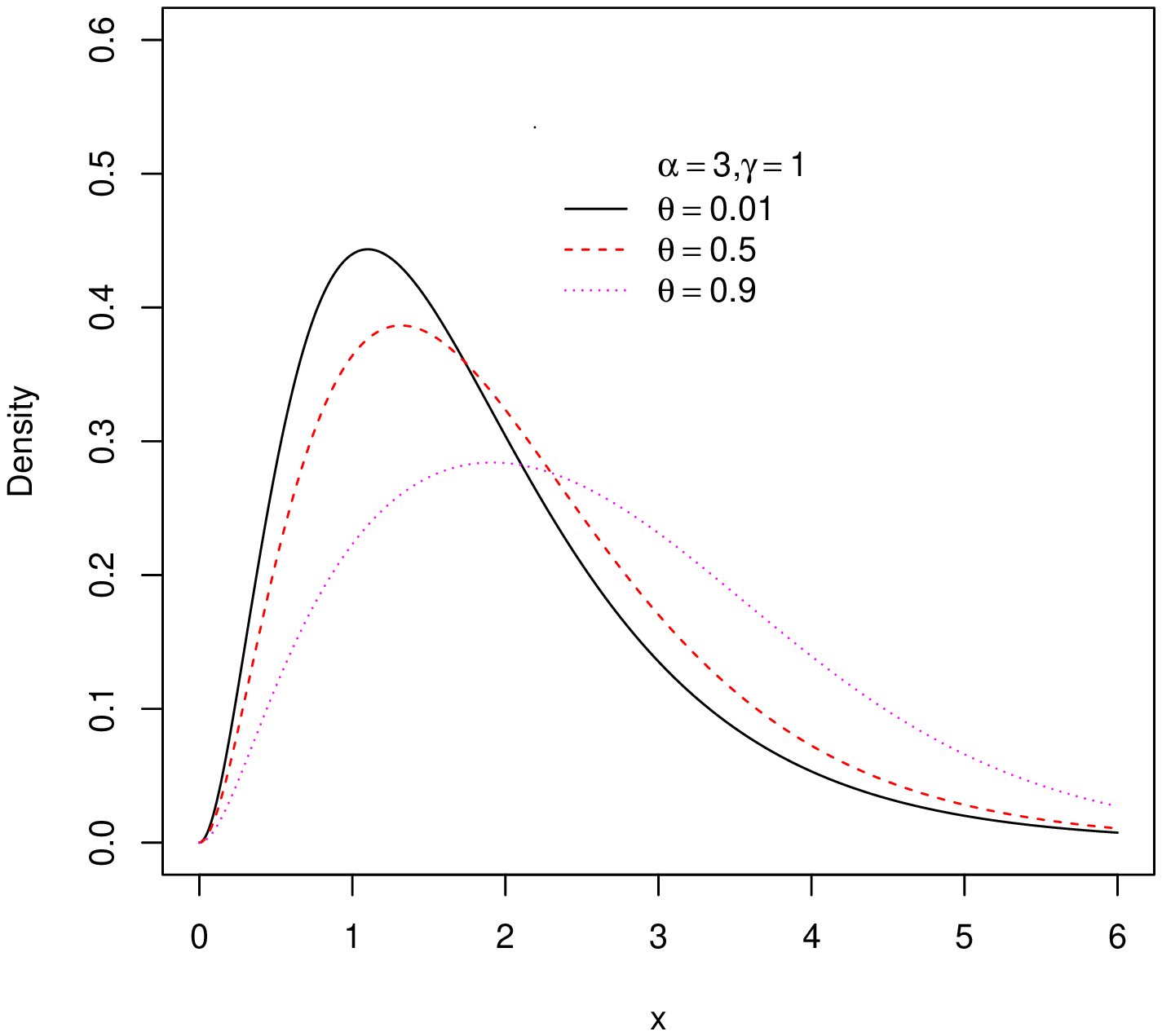}
\includegraphics[scale=0.30]{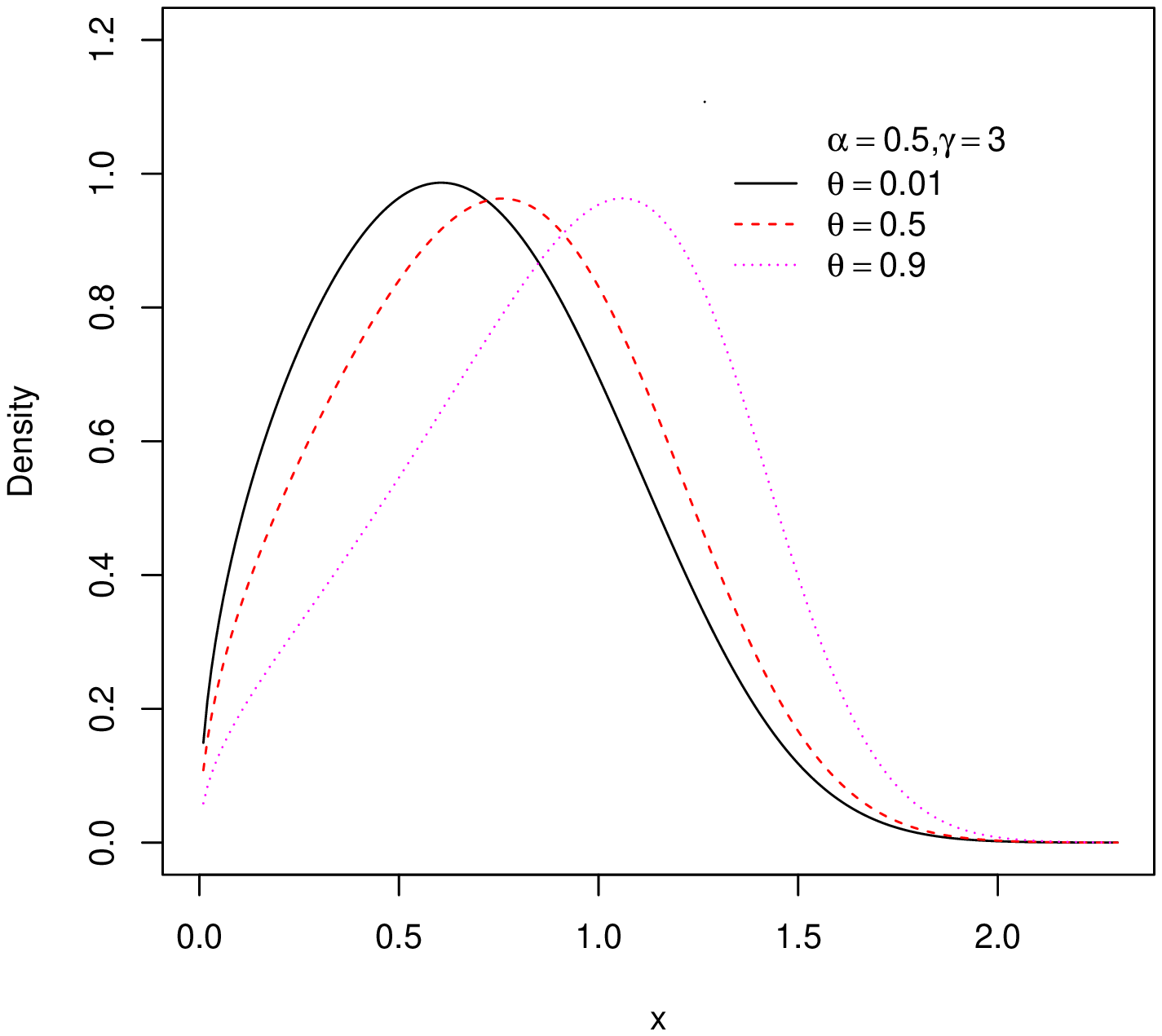}
\includegraphics[scale=0.30]{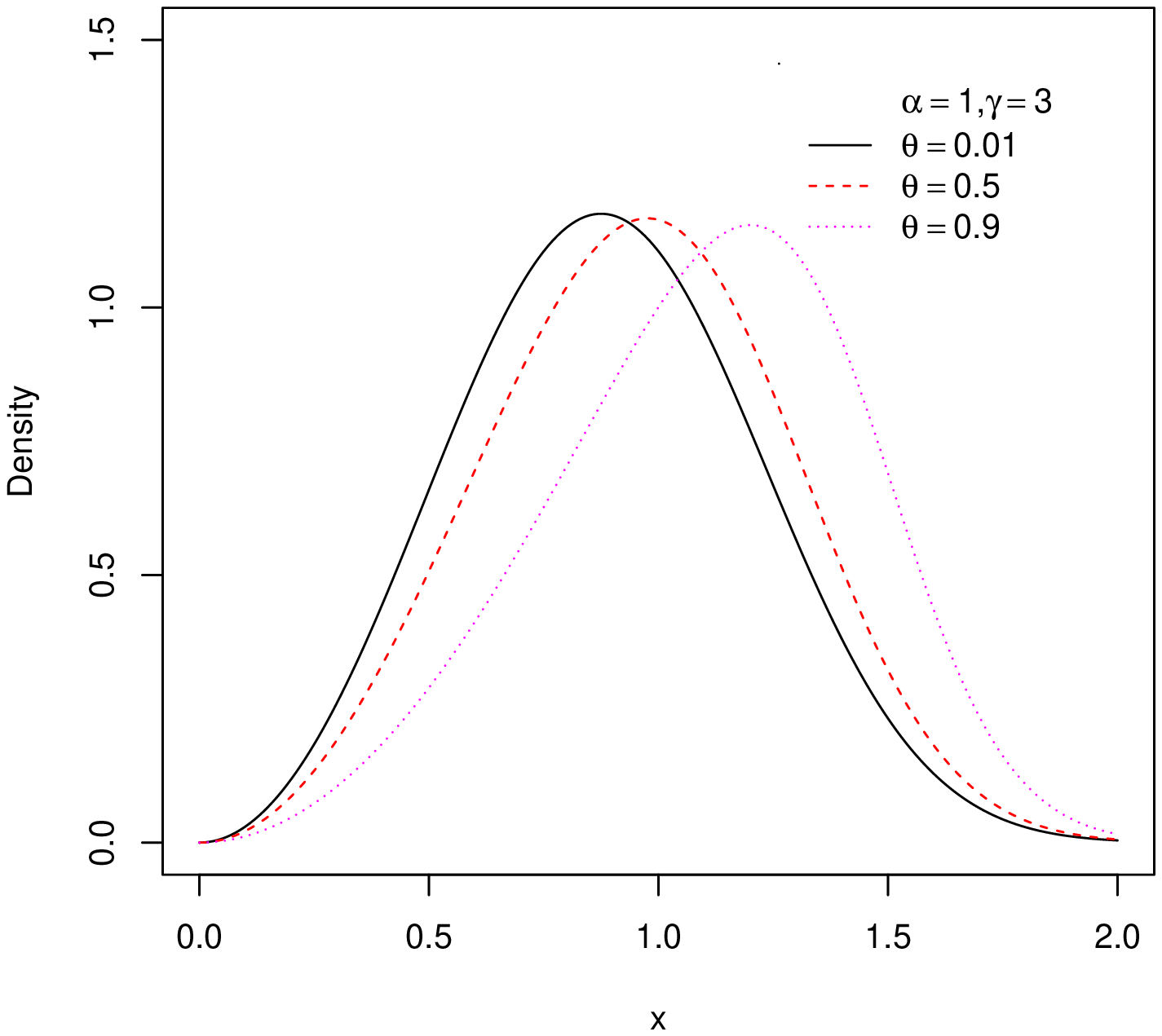}
\includegraphics[scale=0.30]{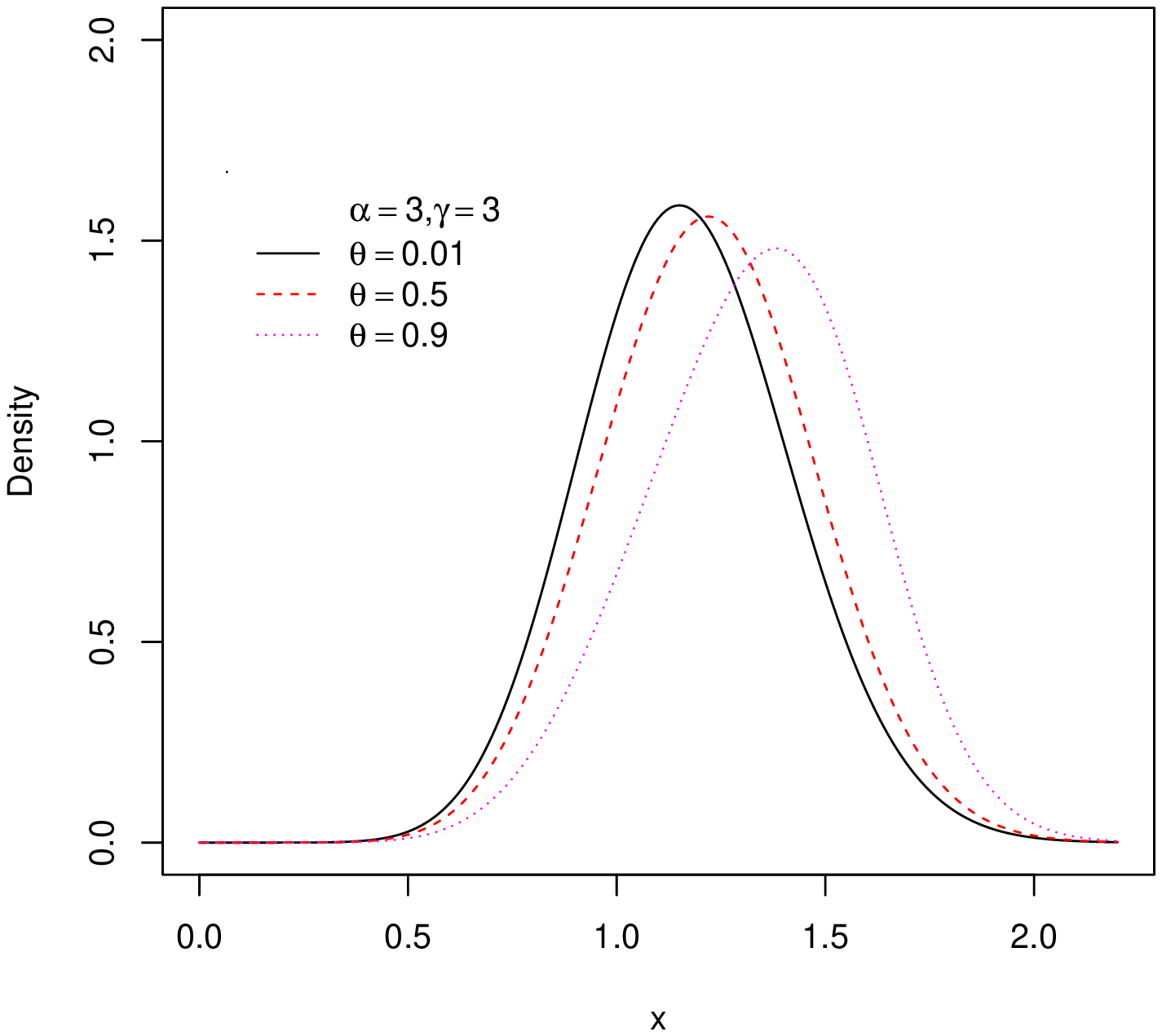}
\caption[]{Plots of probability density function of EWL distribution
for $\beta=1$ and different values $\alpha $, $\gamma $, and $\theta
$.}
\end{figure}

\section{Properties of the distribution}

The probability density function of EWL distribution, which is given
in (\ref{pdf EWL}), tending to zero as $y\rightarrow \infty$. The
EWL leads to EW distribution as $\theta\rightarrow 0$. The following
values of the parameters $\alpha$, $\beta$, $\gamma$ and $\theta$
are of particular interest: (i) $\gamma=1$, EWL reduces to GEL,
which introduced and examined by Mahmoudi and Jafari
\cite{Mahmoudi2011a }; (ii) $\alpha=1$, EWL reduces to CWL; (iii)
$\alpha=\gamma=1$, EWL reduces to CEL, our approach here is
complementary to that of Tahmasbi and Rezaei \cite{Tahmasbi } in the
sense that they consider the distribution $\min(X_1, X_2 . . . ,
X_N)$ while we deal with $\max(X_1, X_2 . . . , X_N)$; (iv)
$\alpha=1$ and $\theta\rightarrow 0$, EWL reduces to Weibull; (v)
$\gamma=1$ and $\theta\rightarrow 0$, EWL reduces to GE.

\subsection{ Quantiles and moments of the EWP distribution }
The $\xi$th quantile of the EWL distribution, which is used for data
generation from the EWL, is given by
\begin{equation}\label{p quantile}
x_{\xi}=\frac{1}{\beta} \big[  -\log(
1-(\frac{1-(1-\theta)^{\xi}}{\theta})^{1/\alpha}) \big]^{1/\gamma}.
\end{equation}
Now we obtain the moment generating function (mgf) and $k$th order
moment of EWL distribution, since some of the most important
features and characteristics such as tending, dispersion, skewness
and kurtosis can be studied through these quantities. For a random
variable $Y$ with EWL distribution the mgf is given by
\begin{equation}
\begin{array}[b]{ll}\label{mgf EWL}
M_{Y}(t)=\frac{‎\alpha\theta}{\log(1-\theta)}\sum^{\infty}_{n=1}\sum^{\infty}_{k=0}\sum^{\infty}_{j=0}(-1)^{j+1}\frac{t^{k}
\theta^{n-1}}{\beta^{k}k!} \Gamma(1+\frac{k}{\gamma}){n\alpha-1
\choose j}(j+1)^{-(1+\frac{k}{\gamma})}.
\end{array}
\end{equation}
$M_{Y}(t)$ in (\ref{mgf EWL}) can be used to obtain the $k$th order
moment of EWL distribution. We have
\begin{equation}\label{meank EWL}
\begin{array}[b]{ll}
E(Y^{k})=\frac{\alpha\theta\Gamma(1+\frac{k}{\gamma})}{\beta^{k}\log(1-‎\theta‎)}\sum^{\infty}_{n=1}\sum^{\infty}_{j=0}(-1)^{j+1}{n\alpha-1
\choose j}\theta^{n-1} (j+1)^{-(1+\frac{k}{\gamma})}.
\end{array}
\end{equation}
\begin{prop}
The random variable $Y$ with pdf given by (\ref{pdf EWL}) has mean
and variance given, respectively, by
\begin{equation}\label{mean EWL}
E(Y)=\frac{\alpha\theta\Gamma (1+\frac{1}{\gamma}
)}{\beta\log(1-‎\theta‎)}\sum^{\infty}_{n=1}\sum^{\infty}_{j=0}(-
1)^{j+1} \theta^{n-1}{n\alpha - 1 \choose j}
(j+1)^{-(1+\frac{1}{\gamma})} ,
\end{equation}
and
\begin{equation}\label{var EWL}
\begin{array}[b]{ll}
Var(Y)=\frac{\alpha\theta\Gamma (1+\frac{2}{\gamma} )}{\beta^{
2}\log(1-‎\theta‎)} \sum^{\infty}_{n=1}\sum^{\infty}_{j=0} (-
1)^{j+1} \theta^{n-1} {n\alpha - 1 \choose j}
(j+1)^{-(1+\frac{2}{\gamma})}-E^2(Y),
\end{array}
\end{equation}
where $E(Y)$ is given in Eq. (\ref{mean EWL}).
\end{prop}
%Note that for
%positive integer values of $\alpha$, the index $j$ in Eqs. (\ref{mgf
%EWL})-(\ref{var EWL}) stops at $n\alpha-1$.
\subsection{The survival, hazard and mean residual life functions}
Using (\ref{cdf EWL}) and (\ref{pdf EWL}), survival function (also
known as reliability function) and hazard function (also known as
failure rate function) of the EWL distribution are given,
respectively, by
\begin{equation}\label{survive}
S(y)=1 -F(y) =1-‎\frac{\log \big(1-    ‎\theta‎
(1-e^{-‎(\beta y‎)^{‎\gamma‎}} ) ^{‎\alpha‎} \big
)}{\log (1-‎\theta‎) }‎ ,~~y>0,
\end{equation}
and
\begin{equation}\label{hazard}
h(y)=‎‎\frac{‎\alpha ‎\theta ‎‎\gamma‎\beta
^{‎\gamma‎} y^{‎\gamma -1‎} e^{-‎(\beta
y‎)^{‎\gamma‎}}  \big(1-e^{-‎(\beta y‎)^{‎\gamma‎}}
\big)^{‎\alpha -1‎} ‎‎‎}{\big( ‎\theta‎
(1-e^{-‎(\beta y‎)^{‎\gamma‎}} ) ^{‎\alpha‎}-1\big )
\big[ \log (1-‎\theta‎)-\log   \big( 1-‎\theta‎
(1-e^{-‎(\beta y‎)^{‎\gamma‎}} ) ^{‎\alpha‎}\big )
\big]}‎
\end{equation}
\begin{prop}
The limiting behavior of hazard function of EWL distribution in
(\ref{hazard}) is\\
(i) for $0<\gamma<1$, $\lim_{y\rightarrow 0}h(y)=\left\{
\begin{array}{lc}
\infty  & 0<\alpha \leq1 \\
0 & \alpha >1, \end{array} \right.$ and $\lim_{y\rightarrow
\infty}h(y)=0.$\\
(ii) for $\gamma=1$, $\lim_{y\rightarrow 0}h(y)=\left\{
\begin{array}{lc}
\infty  & 0<\alpha <1 \\
-\frac{\theta\beta}{\log(1-\theta)} & \alpha =1 \\
0 & \alpha>1, \end{array} \right.$ and $\lim_{y\rightarrow
\infty}h(y)=\beta.$\\
(iii) for $\gamma>1$, $\lim_{y\rightarrow 0}h(y)=0,$ for each value
$\alpha>0$ and $\lim_{y\rightarrow \infty}h(y)=\infty$.
\end{prop}
\begin{pf}
The proof is a forward calculation and is omitted.
\end{pf}
The graphs of hazard rate function of EWL distribution for $\beta=1$
and various values $\alpha$, $\gamma$ and $\theta$ are displayed in
Fig. 2.
\begin{figure}[t]
\centering
\includegraphics[scale=0.30]{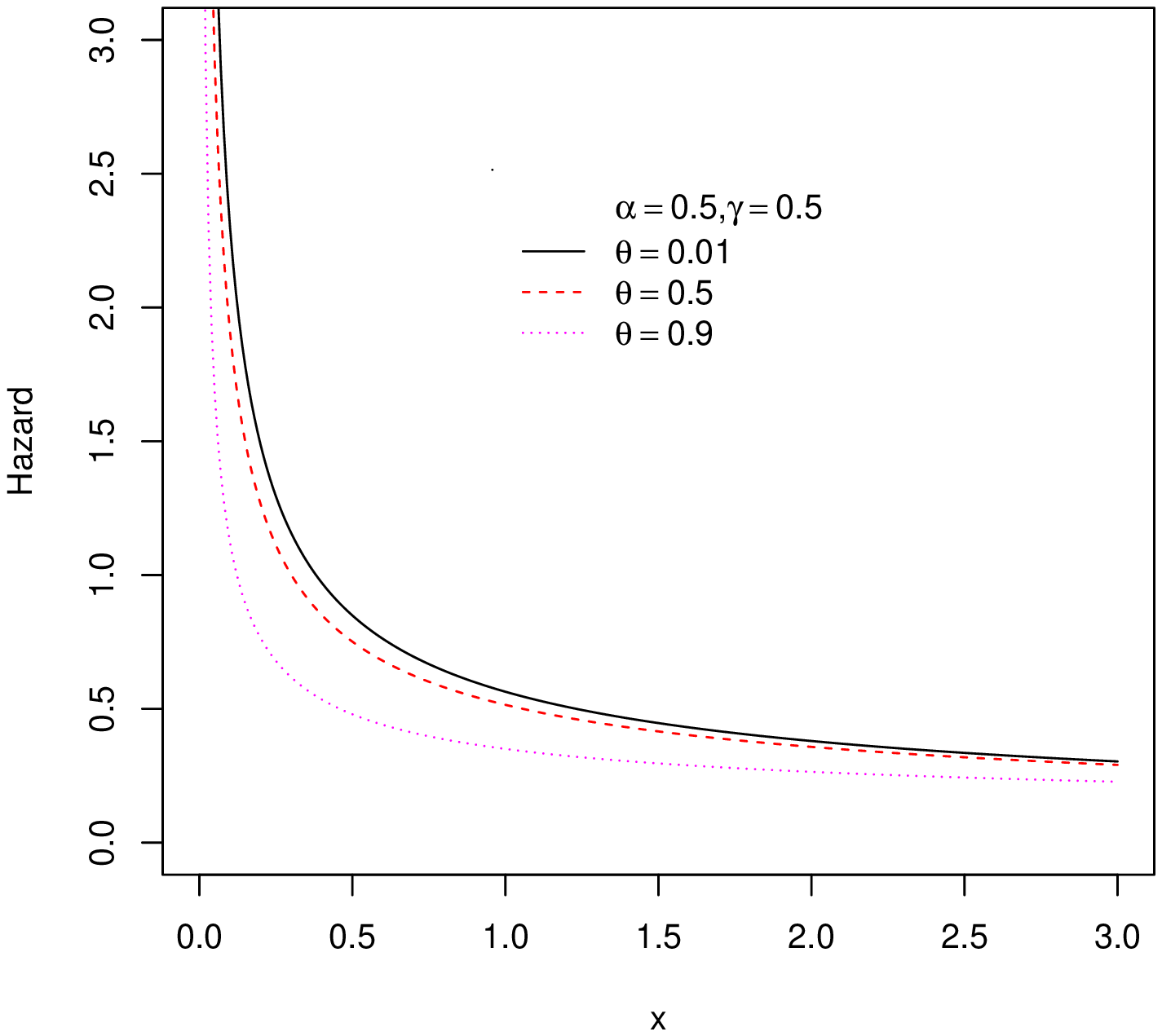}
\includegraphics[scale=0.30]{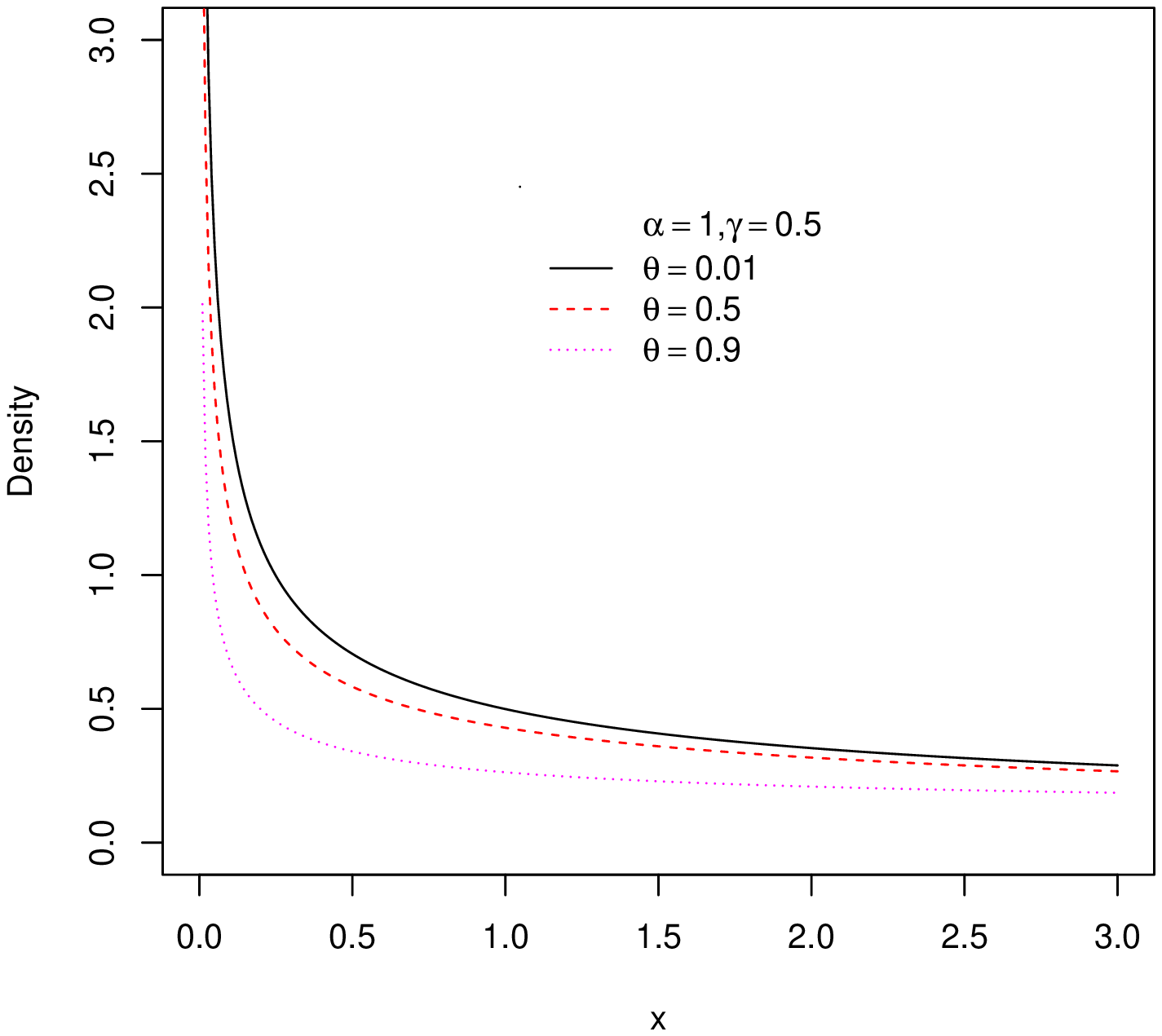}
\includegraphics[scale=0.30]{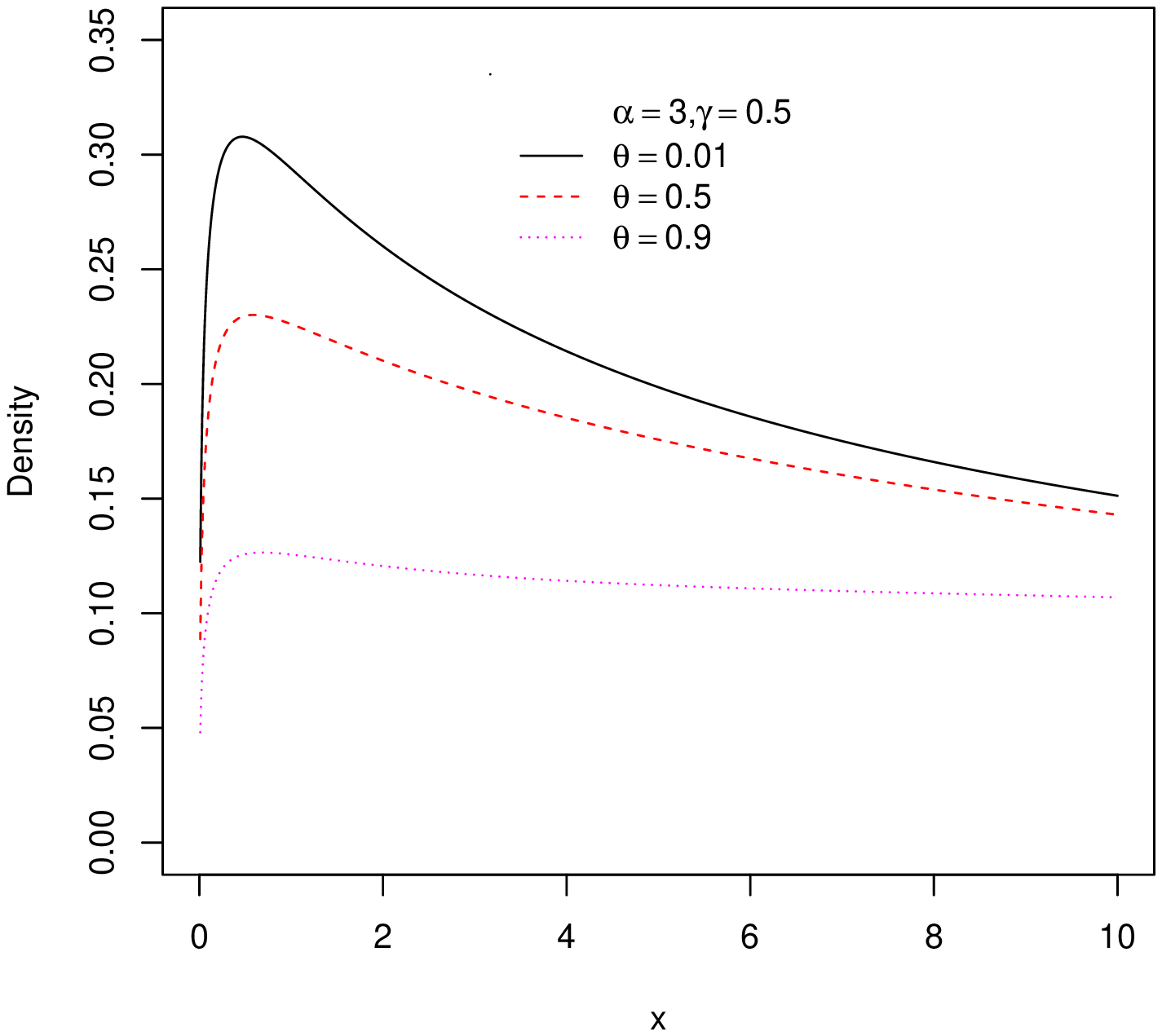}
\includegraphics[scale=0.30]{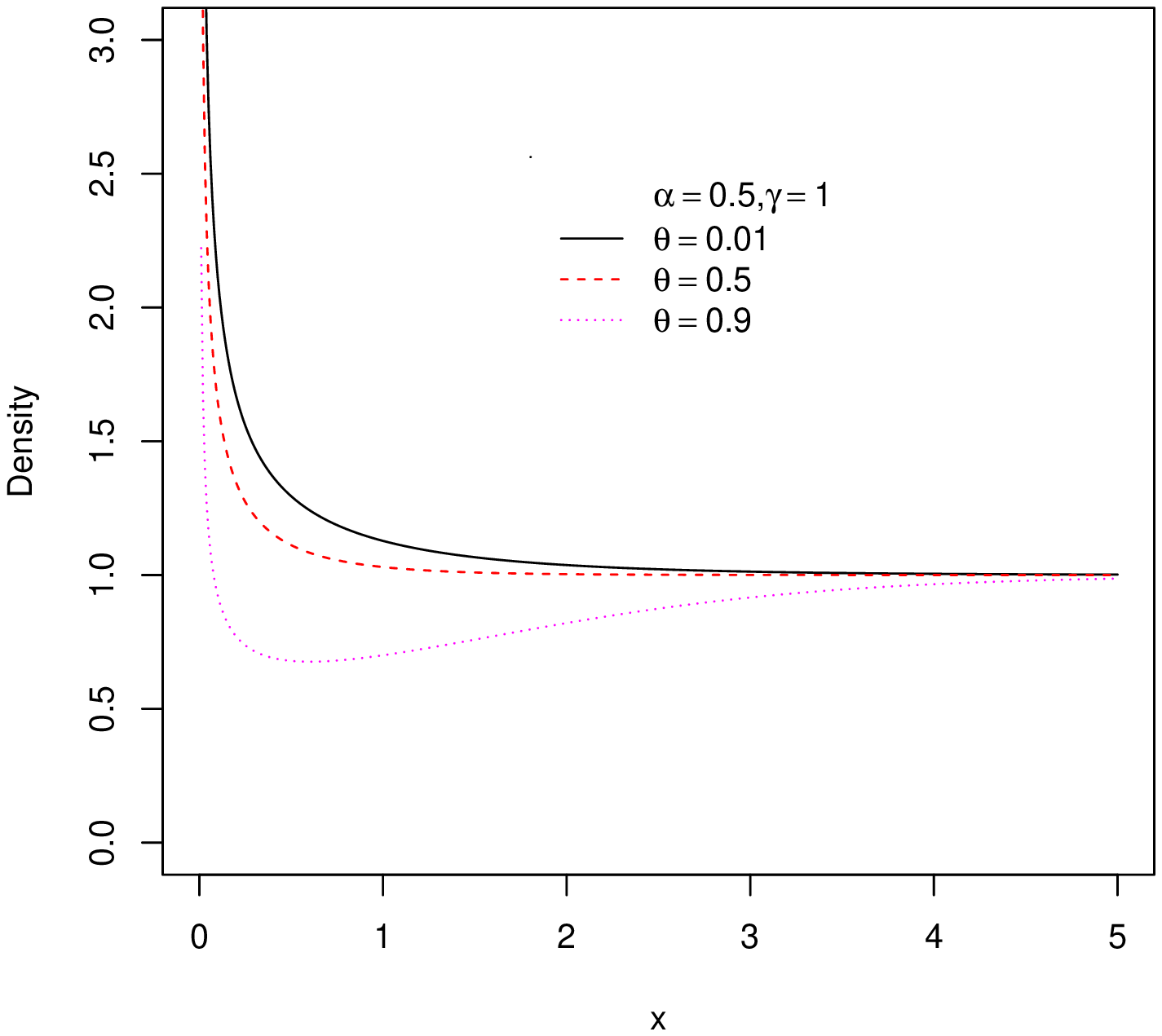}
\includegraphics[scale=0.30]{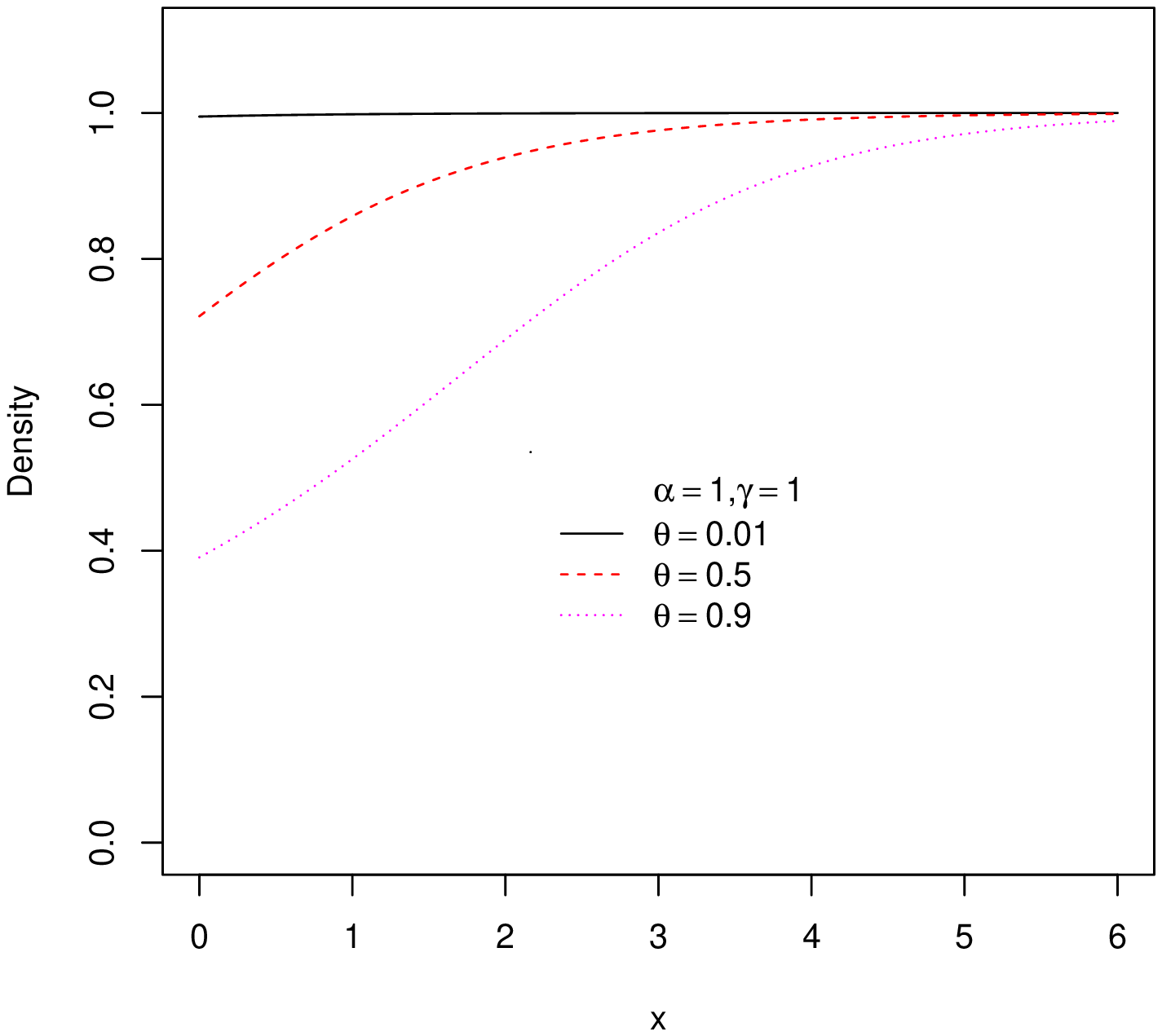}
\includegraphics[scale=0.30]{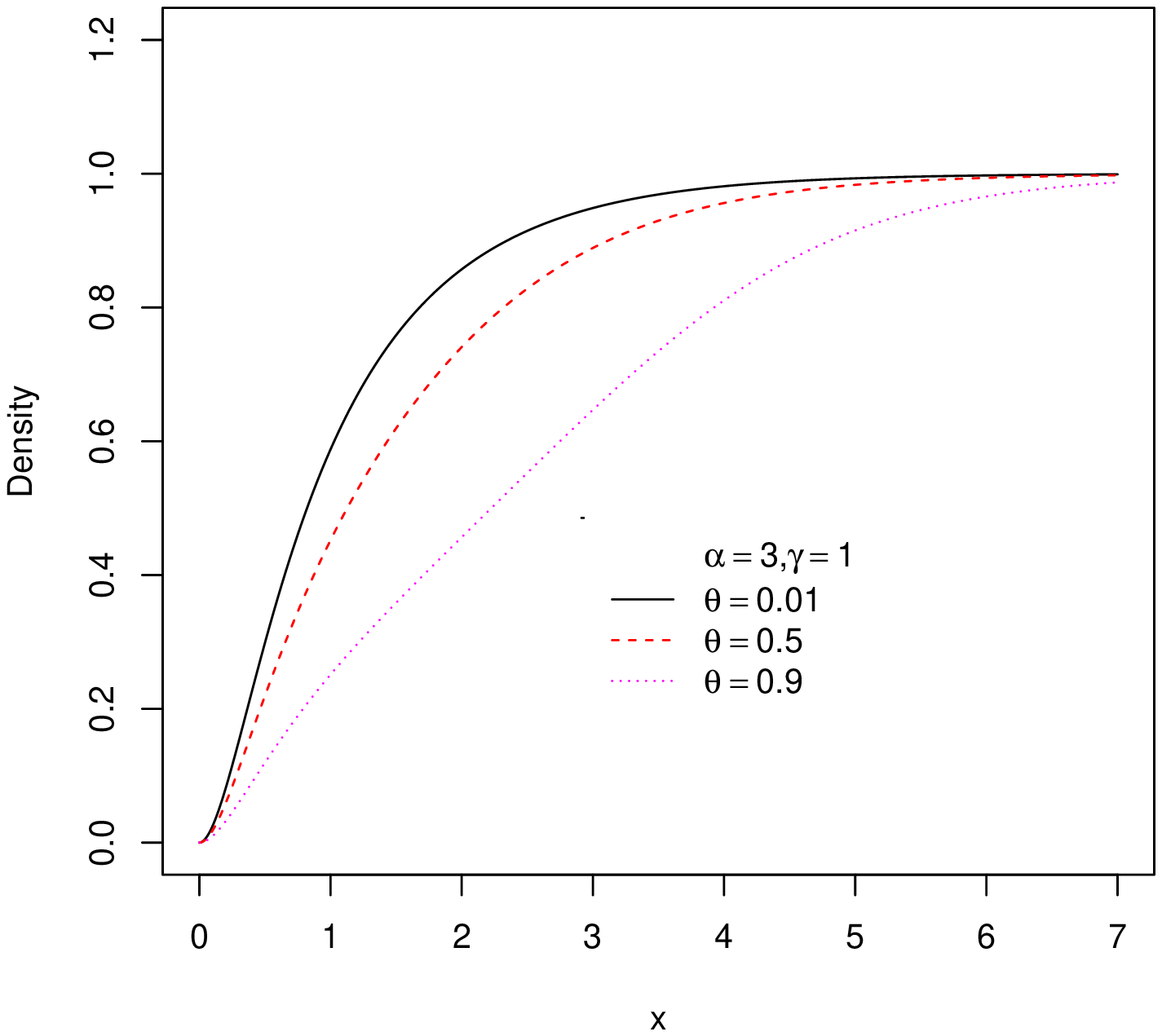}
\includegraphics[scale=0.30]{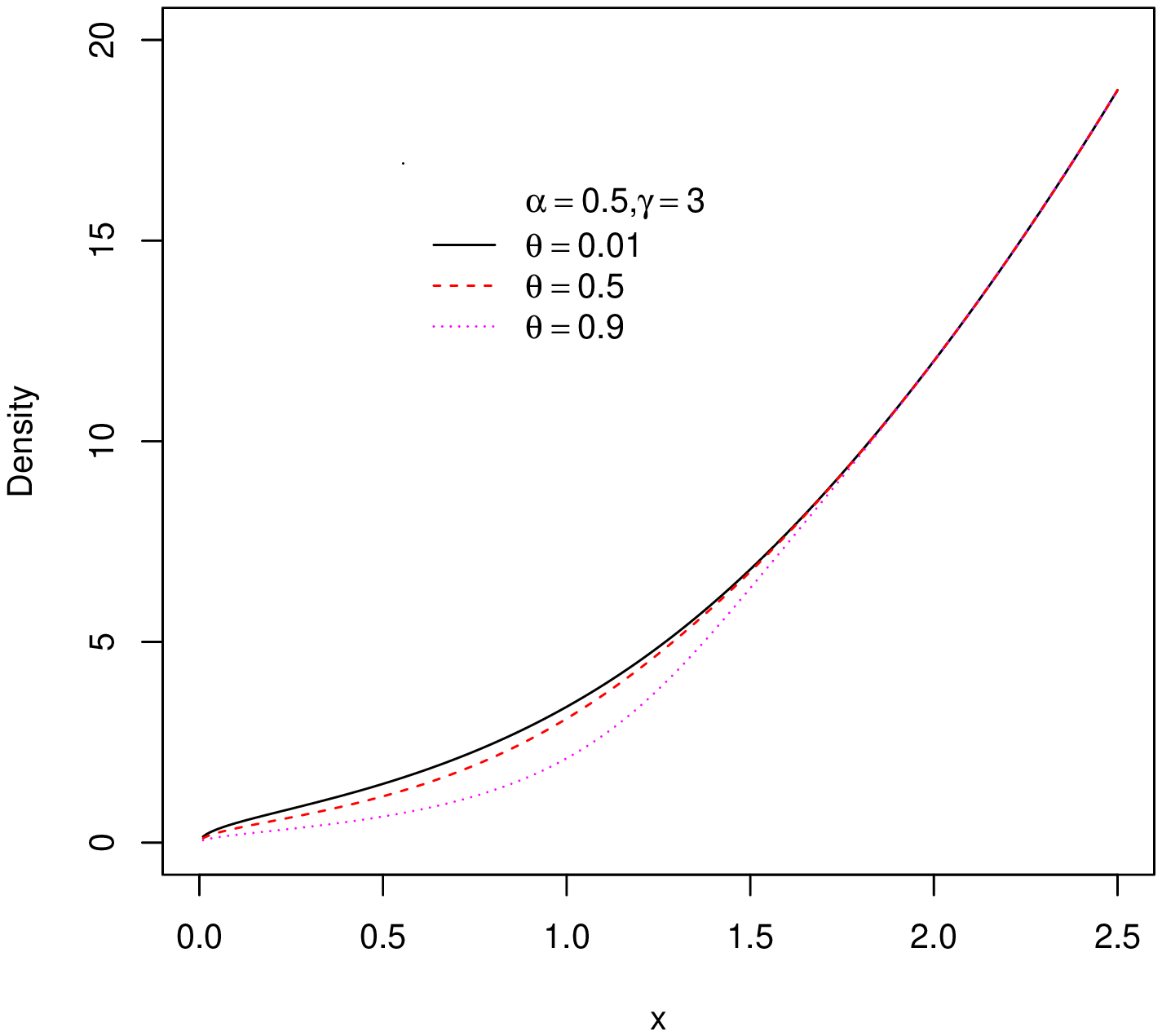}
\includegraphics[scale=0.30]{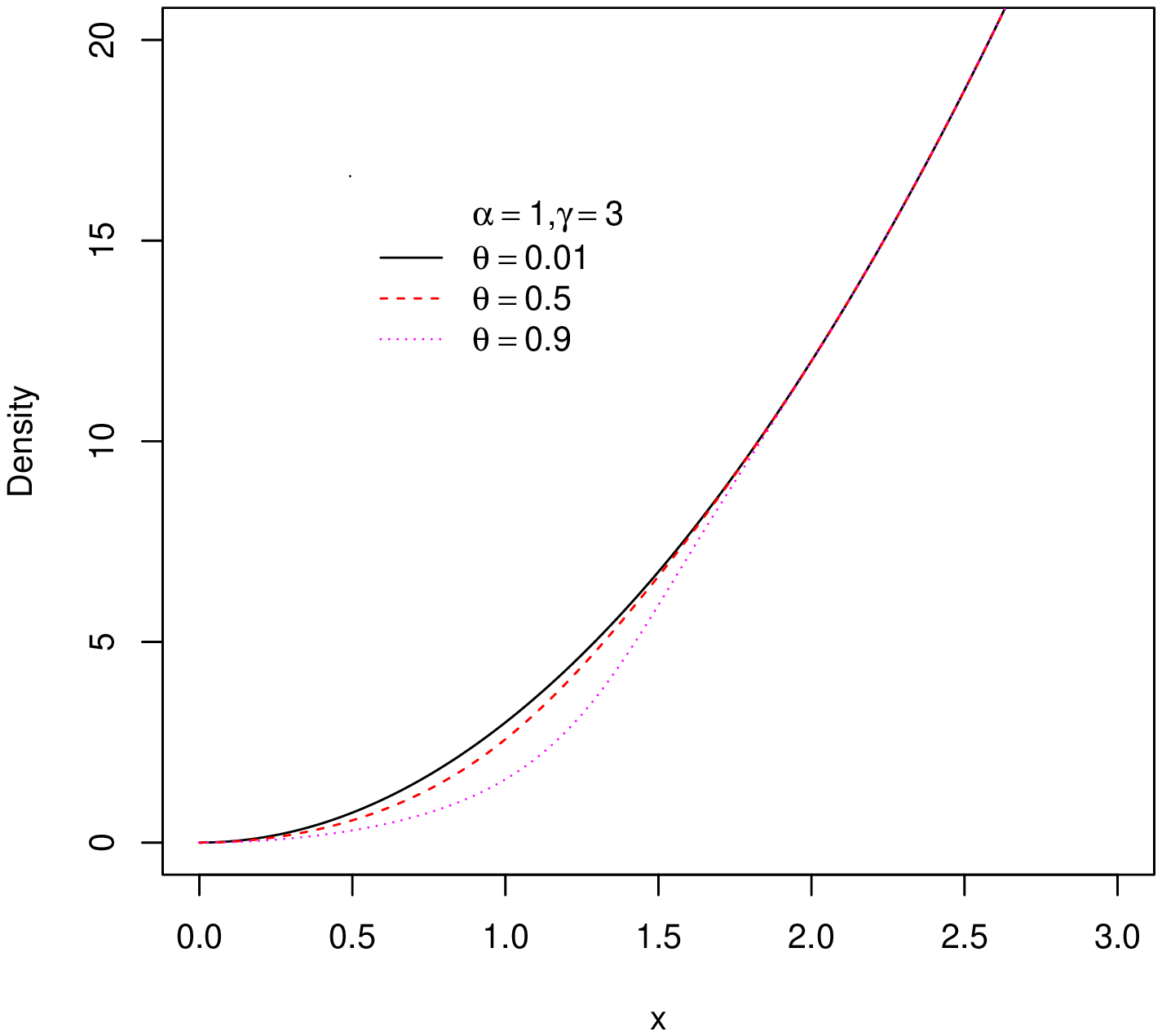}
\includegraphics[scale=0.30]{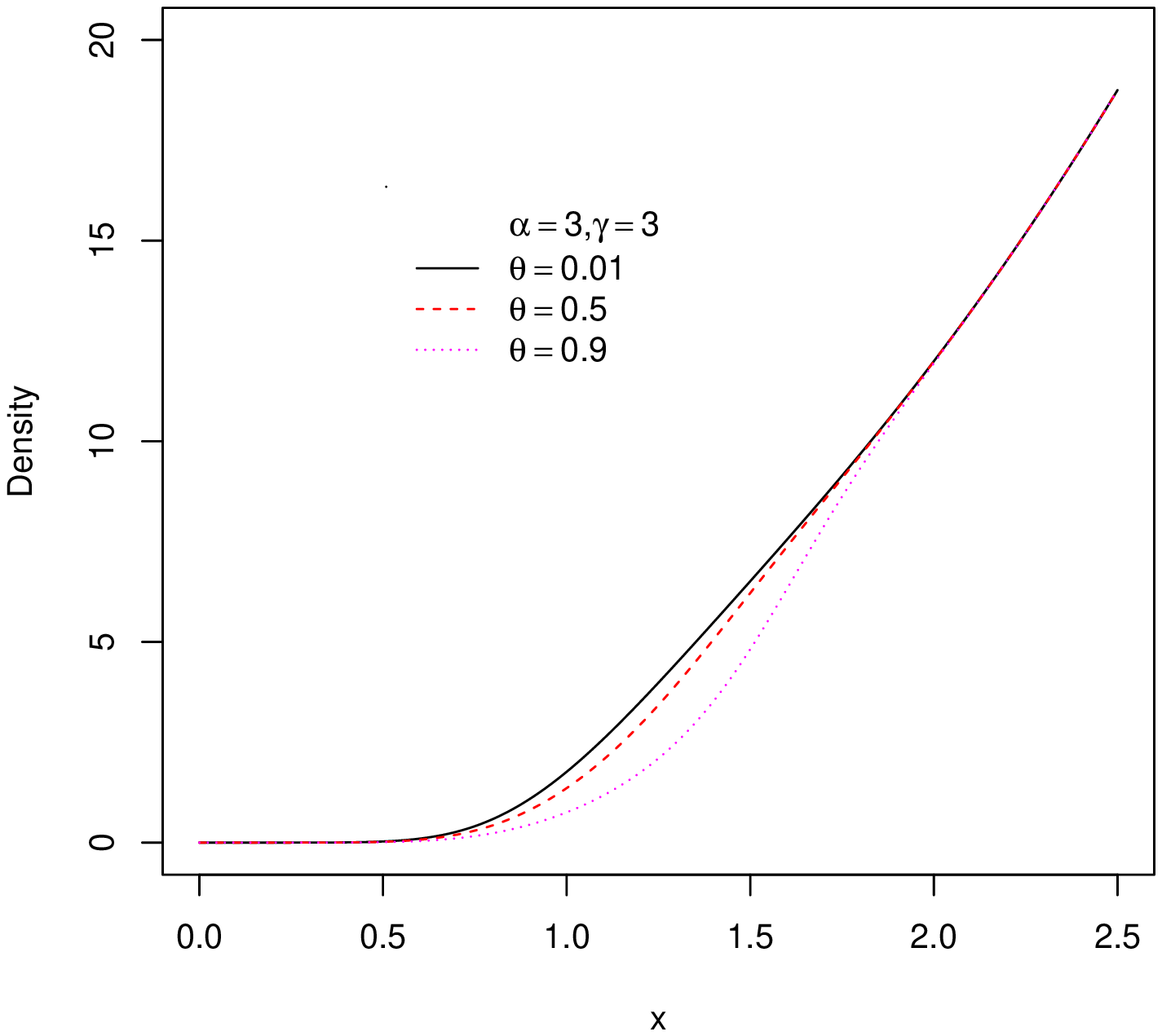}
\caption[]{Plots of hazard rate function of EWL distribution for
$\beta=1$ and different values $\alpha $, $\gamma $, and $\theta $.}
\end{figure}

Given that a component survives up to time $t\geq0$, the residual
life is the period beyond $t$ until the time of failure and defined
by the conditional random variable $X-t|X > t$. The mean residual
life (MRL) function is an important function in survival analysis,
actuarial science, economics and other social sciences and
reliability for characterizing lifetime. In reliability, it is well
known that the MRL function and ratio of two consecutive moments of
residual life determine the distribution uniquely (Gupta and Gupta,
\cite{Gupta1983 }).

In what seen this onwards we use the equations
\begin{equation*}\label{incom int1}
\int_{t}^{\infty}y^{\gamma+s-1}e^{-(k+1)(\beta
x)^{\gamma}}dy=\frac{1}{\gamma\beta^{‎\gamma+s‎}}(k+1)^{-(1+\frac{s}{\gamma})}\Phi(1+\frac{s}{\gamma};(k+1)(\beta
t)^{\gamma}),
\end{equation*}
and
\begin{equation*}\label{incom int2}
\int_{0}^{t}y^{\gamma+s-1}e^{-(k+1)(\beta
x)^{\gamma}}dy=\frac{1}{\gamma\beta^{\gamma
+s}}(k+1)^{-(1+\frac{i}{\gamma})}\gamma(1+\frac{s}{\gamma};(k+1)(\beta
t)^{\gamma}),
\end{equation*}
where $\Phi(s;t)= \int_{t}^{\infty}x^{s-1}e^{-x}dx$ is the upper
incomplete gamma function and $\gamma(s;t)=
\int_{0}^{t}x^{s-1}e^{-x}dx$ is the lower incomplete gamma function.
The $r$th order moment of the residual life of the EWL distribution,
which is obtain via the general formula
\begin{equation*}
m_{r}(t)=E\left[(Y-t)^r|Y>t\right]=\frac{1}{S(t)}\int_{t}^{\infty}(y-t)^{r}f(y)dy,
\end{equation*}
where $S(t)$ is the survival function, is given by
\begin{equation}\label{r res}
\begin{array}[b]{ll}
m_{r}(t)&=‎\frac{‎\alpha ‎\theta ‎‎}{S(t) \log(1-‎\theta‎)}\sum_{i=0}^{r}\sum_{j=0}^{\infty}‎
\sum_{k=0}^{\infty} ‎(-1)^{r+k-i+1 } t^{r-i} ‎\theta^{j‎} \beta^{-i‎} (k+1)‎^{-(1+‎\frac{i}{‎\gamma‎}‎) }‎
{{r}\choose{i}}{{\alpha(j+1)-1}\choose{k}}‎\medskip\\
&~~\times \Phi(1+\frac{i}{\gamma}; (k+1) (\beta t)^{\gamma}),
\end{array}
\end{equation}
where $S(t)$ (the survival function of $Y$) is given in
(\ref{survive}). \\
The MRL function of EWL obtain by setting $r=1$ in Eq. (\ref{r
res}). MRL function as well as failure rate function is very
important since each of them can be used to determine a unique
corresponding life time distribution. Life times can exhibit IMRL
(increasing MRL) or DMRL (decreasing MRL). MRL functions that first
decreases (increases) and then increases (decreases) are usually
called bathtub (upside-down bathtub) shaped, BMRL (UMRL).
% The relationship between the behaviors of the two
%functions of a distribution was studied by many authors such as
%Ghitany (1998), Mi (1995), Park (1985), Shanbhag (1970), and Tang et
%al. (1999). For the EWG distribution the MRL function is given in
%the following theorem.

\begin{thm}
The MRL function of the EWL distribution is given by
\begin{equation}\label{MRL EWG}
\begin{array}[b]{ll}
m_{1}(t)&=\frac{‎\alpha ‎\theta}{S(t)
\beta\log(1-‎\theta‎)}\sum_{j=0}^{\infty}\sum_{k=0}^{\infty}(-1)^{k+1}
‎\theta ^j {{\alpha(j+1)-1}\choose{k}}
(k+1)^{-(1+\frac{1}{\gamma})}‎\medskip\\
&~~\times\Phi(1+‎\frac{1}{‎\gamma‎}‎; (k+1) (‎\beta
t‎)^{‎\gamma‎}) -t.
\end{array}
\end{equation}
\end{thm}
The variance of the residual life of the EWL distribution
can be obtained easily using $m_{2}(t)$ and $m_{1}(t)$.\\
On the other hand, the reversed residual life can be defined as the
conditional random variable $t-X|X \leq t$ which denotes the time
elapsed from the failure of a component given that its life is less
than or equal to t. This random variable may also be called the
inactivity time (or time since failure); for more details one can
see Kundu and Nanda (2010) and Nanda et al., (2003). Using (\ref{pdf
EWL}) and (\ref{survive}), the reversed failure (or reversed hazard)
rate function of the EWL is given by
\begin{equation*}
r(y)=\frac{f(y)}{F(y)}=‎‎\frac{‎\alpha ‎\theta ‎\gamma
‎\beta ^{‎\gamma‎}‎‎‎‎ y^{‎\gamma-1‎}
e^{-(‎\beta y‎)^{‎\gamma‎}}\big(1-e^{-(‎\beta
y‎)^{‎\gamma‎}}\big)^{‎\alpha-1‎}}{\big(
‎\theta(1-e^{-(‎\beta y‎)^{‎\gamma‎}})‎^{‎\alpha‎}
-1\big) \log \big[ 1-‎\theta(1-e^{-(‎\beta
y‎)^{‎\gamma‎}})‎^{‎\alpha‎})‎\big]}‎ ‎‎.
\end{equation*}
The $r$th order moment of the reversed residual life can be obtained
by the well known formula
\begin{equation*}
\mu_{r}(t)=E\left[(t-Y)^r|Y\leq
t\right]=\frac{1}{F(t)}\int_{0}^{t}(t-y)^{r}f(y)dy,
\end{equation*}
hence,
\begin{equation}\label{rev residual}
\begin{array}[b]{ll}
\mu_{r}(t)&=\frac{‎\alpha ‎\theta‎ ‎}{F(t) \log
(1-‎\theta‎)
}\sum_{i=0}^{r}\sum_{j=0}^{\infty}\sum_{k=0}^{\infty}
‎\frac{(-1)^{i+k+1} ‎\theta^{j ‎}
t^{r-i}‎‎}{\beta^{i}(k+1)^{1+‎‎\frac{i}{‎\gamma‎}‎‎}}
{{r}\choose{i}} {{(j+1)‎\alpha -1‎}\choose{k}}
‎\gamma‎(1+‎\frac{i}{‎\gamma‎}; (k+1)(‎\beta
t‎)^{‎\gamma‎}).
\end{array}
\end{equation}
The mean and second moment of the reversed residual life of the EWL
distribution can be obtained by setting $r=1,~2$ in (\ref{rev
residual}). Also, using $\mu_{1}(t)$ and $\mu_{1}(t)$ we obtain the
variance of the reversed residual life of the EWL distribution.

\subsection{ Mean deviations from the mean and median}

The amount of scatter in a population can be measured by the
totality of deviations from the mean and median. The mean deviation
from the mean is a robust statistic, being more resilient to
outliers in a data set than the standard deviation. The mean
deviation from the median is a measure of statistical dispersion. It
is a more robust estimator of scale than the sample variance or
standard deviation.

For a random variable $X$ with pdf $f(x)$, cdf $F(x)$, mean $\mu=
E(X)$ and $M = Median(X)$, the mean deviation about the mean and the
mean deviation about the median are defined by
$$\delta_{1}(X)=\int_{0}^{\infty}|x-\mu|f(x)dx=2\mu F(\mu)-2I(\mu)$$
and
$$\delta_{2}(X)=\int_{0}^{\infty}|x-M|f(x)dx=2M F(M)-M+\mu -
2I(M),$$ respectively, where $I(b)=\int^{b}_{0}xf(x)dx$
\begin{thm}
The Mean deviations function of the EWL distribution are
\begin{equation*}\label{MD EWp}
\begin{array}[b]{ll}
\delta_{1}(X)&=\frac{2}{ \log (1-‎\theta)‎}\Big[\mu \log
\left(1-‎\theta (1-e^{-(‎\beta
‎\mu‎‎)^{‎\gamma‎}})‎^{‎\alpha‎}\right)+\alpha\theta\beta^{-1}\sum_{j=0}^{\infty}\sum_{k=0}^{\infty}(-1)^{k}\frac{
\theta^{j}}{(k+1)^{1+\frac{1}{\gamma}}} {{\alpha(j+1)-1}\choose{k}}
‎ \\&~~~\times‎\gamma(1+\frac{1}{\gamma};(k+1)(\beta
\mu)^{\gamma})\Big],
\end{array}
\end{equation*}
and
\begin{equation*}\label{MD EWp}
\delta_{2}(X)=\mu +\frac{2\alpha\theta}{\beta \log
(1-‎\theta)‎)}\sum_{j=0}^{\infty}\sum_{k=0}^{\infty}(-1)^{k}\frac{\theta^{j}}{(k+1)^{1+\frac{1}{\gamma}}}
{{\alpha(j+1)-1}\choose{k}}
‎‎\gamma(1+\frac{1}{\gamma};(k+1)(\beta M)^{\gamma}),
\end{equation*}
where $\mu$ is given in (\ref{mean EWL}) and $M$ is obtained by
setting $\xi=.5$ in (\ref{p quantile}).
\end{thm}

\subsection{ Bonferroni and Lorenz curves}

The Bonferroni and Lorenz curves and Gini index have many
applications not only in economics to study income and poverty, but
also in other fields like reliability, medicine and insurance. The
most remarkable property of the Bonferroni index is that it
overweights income transfers among the poor, and the weights are
higher the lower the transfers occur on the income distribution.
Hence, it is a good measure of inequality when changes in the living
standards of the poor are concerned. There are many problems
especially in labor economics that fall into this category. Using a
version of the assignment model, we show that the Bonferroni index
can be formulated endogenously within a mechanism featuring
efficient assignment of workers to firms. This formulation is useful
in evaluating the interactions between the distribution of skills
and earnings inequality with a special emphasis on the lower tail of
the earnings distribution. Moreover, it allows us to think about
earnings inequality by separately analyzing the contribution of each
economic parameter.

The Bonferroni curve $B_{F}[F(x)]$ is given by
\begin{equation*}
B_{F}[F(x)]=\frac{1}{\mu F(x)}\int^{x}_{0}u f(u) du,
\end{equation*}
or equivalently given by $B_{F}[p]=\frac{1}{\mu
p}\int^{p}_{0}F^{-1}(t) dt$ where $p = F(x)$ and $F^{-1}(t)= \inf\{x
: F(x) \geq t \}.$ The Bonferroni and Lorenz curves of the EWL
distribution are given, respectively, by
\begin{equation*}\label{B_{F} EWl}
\begin{array}{ll}
B_{F}[F(x)]&= \frac{-\alpha\theta}{\beta\mu  \log\left(
1-‎\theta(1-e^{-(‎\beta
x‎)^{‎\gamma‎}})‎^{‎\alpha‎}\right)}
\sum_{j=0}^{\infty}\sum_{k=0}^{\infty}(-1)^{k}\frac{\theta^{j}}{(k+1)^{1+\frac{1}{\gamma}}}
{{\alpha(j+1)-1}\choose{k}} ‎\\
&~~~\times‎\gamma(1+\frac{1}{\gamma};(k+1)(\beta x)^{\gamma}),
\end{array}
\end{equation*}
and
\begin{equation*}
\begin{array}{ll}
L_{F} [F(x)] &=
B_{F}[F(x)]F(x)=\frac{-\alpha\theta}{\beta\mu\log\left(
1-‎\theta‎\right)}\sum_{j=0}^{\infty}\sum_{k=0}^{\infty}(-1)^{k}
\frac{\theta^{j}}{(k+1)^{1+\frac{1}{\gamma}}}
{{\alpha(j+1)-1}\choose{k}} ‎\\
&~~~\times‎\gamma(1+\frac{1}{\gamma};(k+1)(\beta x)^{\gamma}),
\end{array}
 \end{equation*}
 where $\mu$ is the mean of EWL distribution.\\
The scaled total time on test transform of a distribution function
$F$ is defined by
\begin{equation*}
S_{F}[F(t)]=\frac{1}{\mu}\int^{t}_{0}S(u)du.
\end{equation*}
If $F(t)$ denotes the cdf of EWL  distribution then
\begin{equation*}
S_{F}[F(t)]=\frac{1}{\mu}\big [ t-\frac{1}{\gamma \beta}
\sum_{i=0}^{\infty}\sum_{k=0}^{\infty}\frac{(-1)^{i+1} \theta^{k}
}{k i^{1/\gamma}}{{\alpha k}\choose{i}} \Phi(\frac{1}{\gamma};i
(\beta t)^{\gamma})\big].
\end{equation*}
The cumulative total time can be obtained by using formula
$C_F=\int_{0}^{1}S_{F}[F(t)]f(t)dt$ and the Gini index can be
derived from the relationship $G =1-C_{F}$.

\subsection{ R\'{e}nyi and Shannon entropies of the EWL distribution}

Entropy has been used in various situations in science and
engineering. The entropy of a random variable $Y$ is a measure of
variation of the uncertainty. Statistical entropy is a probabilistic
measure of uncertainty or ignorance about the outcome of a random
experiment, and is a measure of a reduction in that uncertainty.
Numerous entropy and information indices, among them the R\'{e}nyi
entropy, have been developed and used in various disciplines and
contexts. For a random variable with the pdf $f$, the R\'{e}nyi
entropy is defined by
\begin{equation}\label{entropy}
I_{R}(r)=\frac{1}{1-r}\log\left[\int_{\mathbb{R}}f^{r}(y)dy\right],
\end{equation}
for $r>0$ and $r\neq 1$. Using the power series
expansion $(1-z)^{\alpha}=\sum_{0}^{\infty}(-1)^{j}{\alpha \choose
j}z^j$ and change of variable $(\beta y)^\gamma=u$, we have

\begin{equation}
%\begin{array}[b]{l}
\int^{\infty}_{0} f^{r}(y)dy
=\left[\frac{\alpha\theta}{\log(1-\theta)}\right]^{r}
(\gamma\beta)^{r-1}\sum^{\infty}_{j=0}\sum^{\infty}_{k=0}(-1)^{k+r}{\alpha(r+j)
-r\choose{k}}\frac{\theta^{j}\Gamma(r+j)}{j!
\Gamma(r)}\frac{\Gamma\left(\frac{r(\gamma-1)+1}{\gamma}\right)}{(k+r)^{\frac{r(\gamma-1)+1}{\gamma}}}.
%%(\theta)^{j}\Gamma(r+j)}{j! \Gamma(r)}  \sum^{\infty}_{k=0}(-1)^{k}
%{\alpha(r+j) -r\choose{k}}\big (r+k \big)^{- \big(‎\frac{r(‎\gamma-1‎)+1}{‎\gamma‎}\big)}‎\medskip\\
%&~~\times \Gamma \big(\frac{-
%r(‎\gamma-1‎)+1}{‎\gamma‎}‎‎\big).
%\end{array}
\end{equation}
Thus, according to definition (\ref{entropy}), the R\'{e}nyi entropy
of EWL distribution is given by
\begin{equation}
\begin{array}[b]{ll}
I_{R}(r)&=\frac{1}{1-r}\log
\Big[\left(\frac{\alpha\theta}{\log(1-\theta)}\right)^{r}
(\gamma\beta)^{r-1}\sum^{\infty}_{j=0}\sum^{\infty}_{k=0}(-1)^{k+r}{\alpha(r+j)
-r\choose{k}}\frac{\theta^{j}\Gamma(r+j)}{j!
\Gamma(r)}\medskip\\
&~~\times\frac{\Gamma\left(\frac{r(\gamma-1)+1}{\gamma}\right)}{(k+r)^{\frac{r(\gamma-1)+1}{\gamma}}}\Big].
\end{array}
\end{equation}
The Shannon entropy is defined by $E\left[-\log (f(Y))\right]$. This
is a special case derived from  $\lim_{r\rightarrow 1}I_{R}(r)$.

\section{ Estimation and inference}
In this section, we discuss the estimation of the parameters of EWL
distribution. Let $Y_{1},Y_{2},\cdots,Y_{n}$ be a random sample with
observed values $y_{1},y_{2},\cdots,y_{n}$ from a EWL distribution
with parameters $\alpha,\beta,\gamma$ and $\theta$. Let
$\Theta=(\alpha,\beta,\gamma,\theta)^{T}$ be the parameter vector.
The total log-likelihood function is given by
\begin{equation*}
\begin{array}[b]{ll}
l_{n}\equiv l_{n}(y;\Theta)&= n\log ‎\alpha‎+ n \log
‎‎\theta‎ + n\log‎‎\gamma ‎+ n ‎\gamma ‎\log‎‎\beta‎ +‎(\gamma -1) ‎\sum^{n}_{i=1}\log y_i \medskip \\
&~~~-\sum^{n}_{i=1} (‎\beta y_i‎)^{\gamma} +(‎\alpha -1‎)‎\sum^{n}_{i=1} \log (1- e^{-(‎\beta y_i‎)^{‎\gamma‎}})\medskip \\
&~~~-‎\sum^{n}_{i=1} \log \left( ‎\theta‎ (1- e^{-(‎\beta
y_{i})^{‎\gamma‎}})^{‎\alpha‎} -1 \right) - n \log \left(
\log(1-‎\theta‎) \right).
\end{array}
\end{equation*}
The associated score function is given by $U_{n}(\Theta)=(\partial
l_{n}/\partial \alpha,\partial l_{n}/\partial \beta, \partial
l_{n}/\partial \gamma,\partial l_{n}/\partial \theta)^{T}$, where

\begin{equation*}
\begin{array}{ll}
\frac{\partial l_{n}}{\partial
\alpha}&=‎\frac{n}{‎\alpha‎}+\sum^{n}_{i=1}  \log(1-
e^{-(‎\beta y_i‎)^{‎\gamma‎}})- ‎\theta
‎‎\sum^{n}_{i=1} ‎\frac{‎\log(1-
e^{-(‎\beta y_i‎)^{‎\gamma‎}}) (1-e^{-(‎\beta y_i‎)^{‎\gamma‎}})^{‎\alpha‎}‎}
{‎\theta‎(1-e^{-(‎\beta y_i‎)^{‎\gamma‎}})^{‎\alpha‎} -1},\medskip‎ \\
 \frac{\partial l_{n}}{\partial \beta}& =‎\frac{n
 ‎\gamma‎}{‎\beta‎}- ‎\gamma ‎ \beta ^{‎\gamma
-1‎}\sum^{n}_{i=1} y_i^{‎\gamma‎} + (‎\alpha
-1‎)‎‎\gamma ‎\beta ^{‎\gamma-1‎}‎‎ \sum^{n}_{i=1}
‎\frac{y_i^{‎\gamma‎} e^{-(‎\beta y_i‎)^{‎\gamma‎}}}
{1-e^{-(‎\beta y_i‎)^{‎\gamma‎}}} \medskip \\
&~~-‎\theta ‎\gamma ‎\alpha‎ ‎\beta^{‎ \gamma
-1‎}‎‎‎‎‎‎‎‎‎\sum^{n}_{i=1}‎\frac{y_i^{‎\gamma‎}
e^{-(‎\beta y_i‎)^{‎\gamma‎}} (1-e^{-(‎\beta y_i‎)^
{‎\gamma‎}})^{‎\alpha -1‎}}{\theta‎(1-e^{-(‎\beta y_i‎)^{‎\gamma‎}})^{‎\alpha‎} -1}‎ ,\medskip\\
\frac{\partial l_{n}}{\partial \gamma}&=‎\frac{n}{‎\gamma‎} +n
\log ‎\beta +\sum^{n}_{i=1} \log y _i - \sum^{n}_{i=1}
(‎\beta y_i‎)^{‎\gamma‎} \log (‎\beta y_i‎) \medskip \\
& ~~‎‎+ (‎\alpha -1‎) \sum^{n}_{i=1} ‎\frac{ (‎\beta
y_i‎)^{‎\gamma‎} \log (‎\beta y_i‎)‎‎ e^{-(‎\beta
y_i‎)^{‎\gamma‎}}}{1-e^{-(‎\beta y_i‎)^ {‎\gamma‎}}}-
‎\theta ‎\alpha\sum^{n}_{i=1} ‎\frac{(‎\beta
y_i‎)^{‎\gamma‎} \log (‎\beta y_i‎) e^{-(‎\beta
y_i‎)^{‎\gamma‎}} (1-e^{-(‎\beta y_i‎)^
{‎\gamma‎}})^{‎\alpha -1‎}}{\theta‎(1-e^{-(‎\beta y_i‎)^{‎\gamma‎}})^{‎\alpha‎} -1}‎  ,‎‎‎\medskip \\
\frac{\partial l_{n}}{\partial \theta}&= ‎\frac{n}{‎\theta‎}-
\sum^{n}_{i=1}‎\frac{(1-e^{-(‎\beta
y_{i}‎)^{‎\gamma‎}})^{‎\alpha
‎}}{\theta‎(1-e^{-(‎\beta
y_i‎)^{‎\gamma‎}})^{‎\alpha‎} -1}‎  +
‎\frac{n}{(1-‎\theta‎) \log (1-‎\theta‎)}‎‎.
\end{array}
\end{equation*}
The maximum likelihood estimation (MLE) of $\Theta $, say
$\widehat{\Theta }$, is obtained by solving the nonlinear system
$U_n\left(\Theta \right)=\textbf{0}$. The solution of this nonlinear
system of equation has not a closed form. For interval estimation
and hypothesis tests on the model parameters, we require the
information matrix. The $4\times 4$ observed information matrix is
\[I_n\left(\Theta \right)=-\left[ \begin{array}{cccc}
I_{\alpha \alpha } & I_{\alpha \beta } & I_{\alpha \gamma } & I_{\alpha \theta }\\
I_{\alpha \beta } & I_{\beta \beta } & I_{\beta \gamma }& I_{\beta \theta } \\
I_{\alpha \gamma } & I_{\beta \gamma } & I_{\gamma \gamma}& I_{\gamma \theta } \\
I_{\alpha \theta } & I_{\beta \theta } & I_{\gamma \theta}& I_{\theta \theta } \\
\end{array} \right],\] whose elements are given in Appendix.

Applying the usual large sample approximation, MLE of $\Theta $ i.e.
$\widehat{\Theta }$ can be treated as being approximately
$N_4(\Theta ,{J_n(\Theta )}^{-1}{\mathbf )}$, where $J_n\left(\Theta
\right)=E\left[I_n\left(\Theta \right)\right]$. Under conditions
that are fulfilled for parameters in the interior of the parameter
space but not on the boundary, the asymptotic distribution of
$\sqrt{n}(\widehat{\Theta }{\rm -}\Theta {\rm )}$ is $N_4({\mathbf
0},{J(\Theta )}^{-1})$, where $J\left(\Theta \right)={\mathop{\lim
}_{n\to \infty } {n^{-1}I}_n(\Theta )\ }$ is the unit information
matrix. This asymptotic behavior remains valid if $J(\Theta )$ is
replaced by the average sample information matrix evaluated at
$\widehat{\Theta }$, say ${n^{-1}I}_n(\widehat{\Theta })$. The
estimated asymptotic multivariate normal $N_4(\Theta
,{I_n(\widehat{\Theta })}^{-1} )$ distribution of $\widehat{\Theta
}$ can be used to construct approximate confidence intervals for the
parameters and for the hazard rate and survival functions. An
$100(1-\gamma )$ asymptotic confidence interval for each parameter
${\Theta }_{{\rm r}}$ is given by
\[{ACI}_r=({\widehat{\Theta
}}_r-Z_{\frac{\gamma }{2}}\sqrt{{\hat{I}}^{rr}},{\widehat{\Theta
}}_r+Z_{\frac{\gamma }{2}}\sqrt{{\hat{I}}^{rr}}),\] where
${\hat{I}}^{rr}$ is the (\textit{r, r}) diagonal element of
${I_n(\widehat{\Theta })}^{-1}$ for $r=1,~2,~3,~4,$ and
$Z_{\frac{\gamma }{2}}$ is the quantile $1-\gamma /2$ of the
standard normal distribution.

The likelihood ratio (LR) statistic is useful for comparing the EWL
distribution with some of its special sub-models. We consider the
partition $\eta = (\eta_{1}^T, \eta_{1}^T)^T$ of the EWL
distribution, where $\eta_{1}$ is a subset of parameters of interest
and $\eta_{2}$ is a subset of the remaining parameters. The LR
statistic for testing the null hypothesis $H_{0}:~
\eta_{1}=\eta_{1}^{(0)}$ versus the alternative hypothesis $H_{1}:~
\eta_{1}\neq\eta_{1}^{(0)}$ is given by $w = 2\{\ell(\hat{\eta})-
\ell(\tilde{\eta})\}$, where $\tilde{\eta}$ and $\hat{\eta}$ are the
MLEs under the null and alternative hypotheses, respectively. The
statistic $w$ is asymptotically (as $n\rightarrow\infty$)
distributed as $\chi^{2}_{k}$, where $k$ is the dimension of the
subset $\eta_{1}$ of interest.

\subsection{EM Algorithm}
The MLEs of the parameters $\alpha$, $\beta$, $\gamma$ and $\theta$
in previous section must be derived numerically. Newton-Raphson
algorithm is one of the standard methods to determine the MLEs of
the parameters. To employ the algorithm, second derivatives of the
log-likelihood are required for all iterations. The EM algorithm is
a very powerful tool in handling the incomplete data problem
(Dempster et al., \cite{Dempster}; McLachlan and Krishnan,
\cite{McLachlan}).\\
 Let the complete-data be $Y_{1},\cdots,Y_{n}$
with observed values $y_{1},\cdots,y_{n}$ and the hypothetical
random variable $Z_{1},\cdots,Z_{n}$. The joint probability density
function is such that the marginal density of $Y_{1},\cdots,Y_{n}$
is the likelihood of interest. Then, we define a hypothetical
complete-data distribution for each $(Y_{i},Z_{i})~~i=1,\cdots,n,$
with a joint probability density function in the form
\begin{equation}
g(y,z;\Theta)=f(y|z)f(z) = ‎z‎\alpha ‎\gamma‎ ‎\beta
^{‎\gamma ‎} y^{‎\gamma -1‎}‎‎‎‎ e^{-(‎\beta
y‎)^{‎\gamma‎}} (1-e^{-(‎\beta y‎)^{‎\gamma‎}})^{z
‎\alpha -1‎} ‎\frac{-‎\theta ^{z}‎}{z \log
(1-‎\theta‎)}‎ ,
\end{equation}
where $\Theta=(\alpha,\beta,\gamma,\theta)$, $y>0$ and $z\in
\mathbb{N}$.\\
Under the formulation, the E-step of an EM cycle requires the
expectation of $(Z|Y;\Theta^{(r)})$ where
$\Theta^{(r)}=(\alpha^{(r)},\beta^{(r)},\gamma^{(r)},\theta^{(r)})$
is the current estimate (in the $r$th iteration) of $\Theta$.\\
The pdf of $Z$ given $Y$, say $g(z|y)$ is given by
\begin{equation*}
g(z|y)= ‎‎‎‎\frac{g( y,z)}{f(y)} =  \big(‎\theta
(1-e^{-(‎\beta y‎)^{‎\gamma‎}})^{‎\alpha
‎}\big)‎‎‎‎^{z-1} \big( 1-‎\theta (1-e^{-(‎\beta
y‎)^{‎\gamma‎}})^{‎\alpha ‎}‎\big),
\end{equation*}
and its expected value is
\begin{equation*}
\begin{array}[b]{ll}
E[Z|Y=y]&=‎ \big(1-‎\theta (1-e^{-(‎\beta
y‎)^{‎\gamma‎}})^{‎\alpha}‎ \big)^{-1}‎.
\end{array}
\end{equation*}
The EM cycle is completed  with the M-step by using the maximum
likelihood estimation over $\Theta$, with the missing $Z$'s
replaced by  their conditional expectations given above.\\
The log-likelihood for the complete-data is
\begin{equation*}
\begin{array}{ll}
l^{*}_{n}(y_{1},\cdots,y_{n};z_{1},\cdots,z;\Theta)&  ‎\propto ‎ -n \log (1-‎\theta‎) -\log ‎\theta \sum^{n}_{i=1} z_i‎+
n\log \alpha +n \log \gamma +n\gamma \log \beta \medskip \\
& +(\gamma-1)\sum^{n}_{i=1} \log y_i -
‎\beta^{‎\gamma‎}‎\sum^{n}_{i=1} y_i
^{\gamma}+\sum^{n}_{i=1} (z_i\alpha -1) \log(1-e^{-{(\beta y_i
)}^{\gamma}}).
\end{array}
\end{equation*}
The components of the score function $U^{*}_{n}(\Theta)=
(\frac{\partial l^{*}_{n}}{\partial \alpha},\frac{\partial l^{*}_{n}
}{\partial \beta},\frac{\partial l^{*}_{n}}{\partial \gamma
},\frac{\partial l^{*}_{n}}{\partial \theta})^{T}$ are given by
\begin{equation*}
\begin{array}{ll}
\frac{\partial
l^{*}_{n}}{\partial\alpha}&=\frac{n}{\alpha}+ \sum^{n}_{i=1}  z_i \log (1 -e^{-(\beta y_i )^{\gamma}}),\medskip\\
\frac{\partial l^{*}_{n}}{\partial\beta}&=\frac{n\gamma}{\beta}
-\gamma\beta^{\gamma
-1}\sum^{n}_{i=1}y_i^{\gamma}+\gamma\beta^{\gamma -1}\sum^{n}_{i=1}
(z_i\alpha -1)[\frac{y_i^{\gamma}e^{-(\beta y_i)^{\gamma}}}{1- e^{-(\beta y_i)^{\gamma}}}],\medskip\\
\frac{\partial
l^{*}_{n}}{\partial\gamma}&=‎\frac{n}{‎\gamma‎}‎ +n
\log‎\beta +‎ \sum^{n}_{i=1}\log y_i - \sum^{n}_{i=1} (‎ \beta
y_i‎)^{‎\gamma‎} \log(‎\beta y_i‎)+  \sum^{n}_{i=1}(z_i
‎\alpha -1‎) ‎\frac{(\beta y_i)^{‎\gamma‎}
\log(\beta y_i) e^{-( ‎\beta‎y_i)^{‎\gamma‎}}}{1-e^{-(\beta y_i)^{‎\gamma‎}}}‎, \medskip\\
\frac{\partial
l^{*}_{n}}{\partial\theta}&=‎‎\frac{n}{1-‎\theta‎}-‎\frac{\sum^{n}_{i=1}
z_i}{‎\theta‎}  ‎‎‎‎.
\end{array}
\end{equation*}
From a nonlinear system of equations $U^{*}_{n}(\Theta)=\textbf{0}$,
we obtain the iterative procedure of the EM algorithm as
\begin{equation*}
\begin{array}{l}
\hat{\alpha}^{(t+1)}=\frac{-n}{\sum^{n}_{i=1} z_i^{(t)}\log (1- e^{-(\hat{\beta}^{(t)}y_{i} )^{\hat {\gamma}^{(t)}}})},\medskip\\
\frac{n \hat {\gamma}^{(t)}}{\hat{\beta} ^{(t+1)}} -\sum^{n}_{i=1}
\hat{\gamma} ^{(t)}y_i^{\hat{\gamma}^{(t)}}
(\hat{\beta}^{(t+1)})^{\hat {\gamma}^{(t)}-1}+\sum^{n}_{i=1}
(z_{i}^{(t)} \hat{\alpha}^{(t)} -1) \frac{\hat{\gamma
}^{(t)}y_i^{\hat{\gamma}^{(t)}} ( \hat {\beta}^{(t+1)})
^{\hat{\gamma}^{(t)}-1} e^{-( \hat {\beta
}^{(t+1)}y_i)^{\hat{\gamma}^{(t)}}}}{1-e^{ -(
\hat{\beta}^{(t+1)}y_i)^{\hat{\gamma}^{(t)}}}}
=0,\medskip\\
\frac{n}{\hat{\gamma}^{(t+1)}}+n\log \hat{\beta}^{(t)}+
\sum^{n}_{i=1}\log y_{i}-\sum^{n}_{i=1}\log(\hat{\beta}^{(t)} y_i)(\hat{\beta}^{(t)}y_i)^{\hat{\gamma}^{(t+1)}} \medskip \\
+\sum^{n}_{i=1}(z_i^{(t)} \hat {\alpha} ^{(t)}-1) \frac{( \hat{\beta
}^{(t)}y_i)^{ \hat{\gamma} ^{(t+1)}} \log (\hat{\beta }^{(t)}y_i)
e^{-(\hat {\beta} ^{( t)}y_i)^{\hat{\gamma}
^{(t+1)}}}}{1-e^{-( \hat{\beta}^{(t)}y_i)^{\hat{\gamma}^{(t+1)}} }}=0,\medskip\\
\hat{‎\theta‎} ^{(t+1)}
=‎\frac{\bar{z}^{(t)}}{1+\bar{z}^{(t)}}‎
\end{array}
\end{equation*}
where $\hat\alpha^{(t+1)}$, $\hat\beta^{(t+1)}$ and
$\hat\theta^{(t+1)}$ are found numerically. Hence, for
$i=1,\cdots,n$, we have
\begin{equation*}
z^{(t)}_{i}=\big[ 1-‎‎\hat{‎\theta‎}‎^{(t)}\big(
1-e^{-(‎\hat{‎\beta‎}^{(t) }
y_{i})^{\hat{‎\gamma‎}^{(t)}}}\big)^{\hat{‎\alpha‎}^{(t)}}‎\big]^{-1}.
\end{equation*}

\section{Sub-models of the EWL distribution }

The EWL distribution contains some sub-models for the special values
of parameters $\alpha$, $\beta$ and $\gamma$. Some of these
distributions are discussed here in details.

\subsection{The CWL distribution}
The CWL distribution is a special case of the EWL distribution for
$\alpha=1$. The pdf, cdf and hazard rate function of the CWL
distribution are given, respectively by
\begin{equation}\label{pdf CWP}
f(x)=\frac{\theta ‎\gamma ‎\beta^{‎\gamma‎} x^{‎\gamma
-1‎} e^{-(‎\beta x‎)^{‎\gamma‎}}}{\log (1-‎\theta‎)
\left( ‎\theta (1-e^{-(‎\beta
x)^{‎\gamma‎}‎})‎-1\right)},
\end{equation}
\begin{equation}\label{cdf CWP}
F(x)=‎‎\frac{\log \left(  1-‎\theta (1-e^{-(‎‎\beta
x‎‎)^{‎\gamma‎}})‎\right)}{\log (1-‎\theta‎)}‎,
\end{equation}
and
\begin{equation}\label{hazard CWP}
h(x)=\frac{‎‎\theta ‎\beta ^{‎\gamma‎} x^{‎\gamma -1‎}
e^{-‎(\beta x)^{‎\gamma‎}}} {\left( ‎\theta‎
(1-e^{-‎(\beta x‎)^{‎\gamma‎}} ) -1\right ) \left[ \log
(1-‎\theta‎)-\log \left( 1-‎\theta‎ (1-e^{-‎(\beta
x)^{‎\gamma‎}} ) \right) \right]}.
\end{equation}
According to Eq. (\ref{mean EWL}) the mean of the CWL distribution
is given by
\begin{equation*}
E(X)=‎\frac{\theta \Gamma(1+‎\frac{1}{‎\gamma‎}‎) }{\beta
\log(1-‎\theta‎)} ‎\sum^{\infty}_{n=1}\sum^{n-1}_{j=0}
‎(-1)^{j+1} ‎\theta^{n-1}‎ ‎{n‎-1 \choose{j}‎}
(j+1)^{-‎(\frac{1}{‎\gamma‎}‎ +1)}.
\end{equation*}
One can obtain the Weibull distribution from the CWL distribution by
taking $\theta$ to be close to zero, i.e., $\lim_{\theta \rightarrow
0^{+}}f_{CWL}(x)=f_{W}(x).$

\subsection{The GEL distribution  }

The GEL distribution is a special case of EWL distribution, obtain
by putting $\gamma=1$. This distribution is introduced and analyzed
by Mahmoudi and Jafari \cite{Mahmoudi2011a }. The pdf, cdf and
hazard rate function of the GEL distribution are given, respectively
by
\begin{equation}\label{pdf GEP}
f(x)=\frac{\alpha ‎\theta ‎\beta e^{-‎\beta x‎}
\left(1-e^{-\beta x‎}\right)^{‎\alpha -1‎} ‎}{\log
(1-‎\theta‎) \left(‎\theta(1-e^{-\beta
x}‎)‎^{‎\alpha‎} -1 \right)},
\end{equation}
\begin{equation}\label{cdf GEP}
F(x)=‎\frac{\log \left[ 1-‎\theta(1-e^{-(‎\beta
x‎)^{‎\gamma‎}})^{‎\alpha‎}‎
\right]}{‎\log(1-‎\theta)‎}‎,
\end{equation}
and
\begin{equation}\label{hazard GEP}
h(x)=‎‎\frac{‎\alpha ‎\theta ‎\beta   e^{-‎\beta x}
\big(1-e^{-‎\beta x} \big)^{‎\alpha -1‎} ‎‎‎}{\big(
‎\theta‎ (1-e^{-‎\beta x} ) ^{‎\alpha‎}-1\big ) \big[ \log
(1-‎\theta‎)-\log   \big( 1-‎\theta‎ (1-e^{-‎\beta x} )
^{‎\alpha‎}\big )  \big]}‎‎‎‎‎.
\end{equation}
According to Eq. (\ref{mean EWL}), the mean of the GEL distribution
is given by
\begin{equation*}
E(X)=‎\frac{‎\alpha ‎\theta ‎‎‎}{\beta\log
(1-‎\theta‎)} ‎\sum^{\infty}_{n=1}\sum^{\infty}_{j=0}
‎(-1)^{j+1} ‎\theta^{n-1}‎‎ ‎{n‎\alpha‎‎-1
\choose{j}‎} (j+1)^{-2}‎ ,
\end{equation*}

\subsection{The CEL distribution }
The CEL distribution is a special case of the EWL distribution for
$\gamma=\alpha=1$. Our approach here is complementary to that of
Tahmasbi and Rezaei \cite{Tahmasbi } in the sense that they consider
the distribution $Y=\min(X_1, X_2 . . . , X_N)$ while we deal with
$Y=\max(X_1, X_2 . . . , X_N)$. \newline
The pdf, cdf and hazard rate
function of the CEL distribution are given, respectively by
\begin{equation}\label{pdf CEP}
f(x)=\frac{\theta ‎\beta e^{-‎\beta x‎} }{\log (1-‎\theta)
\left( ‎\theta(1-e^{-‎\beta x‎})-1
\right)‎‎}‎‎‎‎‎,
\end{equation}
\begin{equation}\label{cdf CEP}
F(x)=‎\frac{\log \left(   1-‎\theta (1-e^{-‎\beta
x‎})‎\right)}{\log (1-‎\theta‎)}‎‎,
\end{equation}
and
\begin{equation}\label{hazard CEP}
h(x)=\frac{\theta \beta  e^{-\beta x}}{\left( ‎\theta‎
(1-e^{-‎\beta x} ) -1\right) \left[ \log (1-‎\theta‎)-\log
\left( 1-‎\theta‎ (1-e^{-‎\beta x} ) \right ) \right]}.
\end{equation}
According to Eq. (\ref{mean EWL}), the mean of CEL distribution is
given by
\begin{equation*}
E(X)=\frac{\theta}{\beta\log (1-\theta)}
\sum^{\infty}_{n=1}\sum^{n-1}_{j=0}(-1)^{j+1}\theta^{n-1}{n-1
\choose{j}}(j+1)^{-2}.
\end{equation*}

\section{Applications of EWL distribution to lifetime data}

To show the superiority of the EWG distribution, we compare the
results of fitting this distribution to some of theirs sub-models
such as WG, EW, GE and Weibull distributions, using two real data
sets. The required numerical evaluations are implemented using the R
softwares. The empirical scaled TTT transform (Aarset,
\cite{Aarset}) and Kaplan-Meier curve can be used to identify the
shape of the hazard function.
%The scaled TTT transform is convex (concave) if the
%hazard rate is decreasing ( increasing), and for bathtub (unimodal)
%hazard rates, the scaled TTT transform is first convex (concave) and
%then concave (convex).
The first data set is given by Birnbaum and Saunders (1969) on the
fatigue life of 6061-T6 aluminum coupons cut parallel with the
direction of rolling and oscillated at 18 cycles per second. The
data set consists of 101 observations with maximum stress per cycle
31,000 psi.

%\begin{figure}
%\centering
%\includegraphics[scale=0.45]{TTT1.eps}
%\includegraphics[scale=0.45]{kaplan1.eps}
%\caption[]{TTT plots and Kaplan-Meier curves of data sets 1 and 2.}
%\end{figure}\\

The TTT plot and Kaplan-Meier curve for two series data in Fig. 3
shows an increasing hazard rate function and, therefore, indicates
that appropriateness of the EWG distribution to fit these data.
Table 1 lists the MLEs of the parameters, the values of K-S
(Kolmogorov-Smirnov) statistic with its respective \textit{p}-value,
-2log(L), AIC (Akaike Information Criterion), AD (Anderson-Darling
statistic) and CM (Cramer-von Mises statistic) for the first data.
These values show that the EWG distribution provide a better fit
than the WG, EW, GE and Weibull for fitting the first data.

 We apply the Arderson-Darling (AD) and Cramer-von Mises (CM) statistics, in order to verify which
 distribution fits better to this data. The AD and CM test statistics are described in details in
Chen and Balakrishnan \cite{Chen}. In general, the smaller the
values of AD and CM, the better the fit to the data. According to
these statistics in Table 1, the EWG distribution fit the first data
set better than the others.

\begin{table}[htp!]
\centering \caption{MLEs(stds.), K-S statistics, \textit{p}-values,
$-2\log(L)$ and AIC for the strengths of 1.5 cm glass fibres.}
\begin{small}
\begin{tabular}{|l|lcccccc|}
\hline
Dist.& MLEs(stds.) & K-S  & \textit{p}-value &$-2\log(L)$& AIC& AD& CM  \\
\hline EWL & $\begin{array}{l}
 \hat{\alpha}= 5.1498 (6.3054),
\hat{\beta}=0.0096 (0.0025)\\
\hat{\gamma}=3.0535 (1.3559), \hat{\theta}=0.1383 (1.1994)
\end{array}$
&0.0707&  0.6942 & 913.204 &921.204&0.378&0.141 \\
EW & $\begin{array}{l}
 \hat{\alpha}= 9.2(3.759),
\hat{\beta}=0.011(0.001)\\
\hat{\gamma}=2.405(0.335)
\end{array}$
& 0.0853 &0.4536&914.068 & 920.068 & 0.568& 0.178 \\
%\hline
GE &~ $\hat{\alpha}$=279.938(1.04), $\hat{\beta}$= 0.045(3.186)&0.1088   & 0.1825  &925.668 &929.668&1.393& 0.308  \\
%\hline
GEL &~$\begin{array}{l}
 \hat{\alpha}=  649.481(498.428),
\hat{\beta}=0.0588(0.01)\\
 \hat{\theta}=0.879 (0.1709)
\end{array}$
 &0.1114&   0.1628 &923.711&929.711 &1.362& 0.314 \\
%\hline
WL &~ $\begin{array}{l} \hat{\beta}= 0.007 (0.0004),
\hat{\gamma}=5.55( 0.926)\\
\hat{\theta}=0.11( 0.94)
\end{array}$
&0.0991 & 0.2749&  927.630 &933.630&1.835 & 0.341 \\
%\hline
Weibull &~  $\hat{\beta}$= 0.0069(0.00012), $\hat{\gamma}$=5.99( 0.452)&0.1005   & 0.2595& 926.557 &930.557&1.382& 0.276\\
\hline
\end{tabular}
\end{small}
\end{table}

Using the likelihood ratio (LR) test, we test the null hypothesis
H0: WG versus the alternative hypothesis H1: EWG, or equivalently,
H0: $\alpha=1$ versus H1: $\alpha\neq 1$. The value of the LR test
statistic and the corresponding \textit{p}-value are 3.706 and
0.0542, respectively. Therefore, the null hypothesis (WG model) is
rejected in favor of the alternative hypothesis (EWG model) for a
significance level $>$ 0.0542. For test the null hypothesis H0: EW
versus the alternative hypothesis H1: EWG, or equivalently, H0:
$\theta=0$ versus H1: $\theta\neq 0$, the value of the LR test
statistic is 13.284 (\textit{p}-value = 0.0003), which includes that
the null hypothesis (EW model) is rejected in favor of the
alternative hypothesis (EWG model) for a significance level $>$
0.0003. We also test the null hypothesis H0: Weibull versus the
alternative hypothesis H1: EWG, or equivalently, H0:
$(\alpha,\theta)=(1,0)$ versus H1: $(\alpha,\theta)\neq(1,0)$. The
value of the LR test statistic is 22.712 (\textit{p}-value =
0.0000), which includes that the null hypothesis (Weibull model) is
rejected in favor of the alternative hypothesis (EWG model) for any
significance level.
The second data represent the strength data measured in GPA, for
single carbon fibers and impregnated 1000-carbon fiber tows. Single
fibers were tested under tension at gauge lengths of 1, 10, 20 and
50 mm. Impregnated tows of 1000 fibers were tested at gauge lengths
of 20, 50, 150 and 300 mm. For illustrative purpose, we will be
considering the single fibers of 10 mm in gauge length, with sample
sizes $n$=63. This data was reported by Badar and Priest
\cite{Badar}.

The MLEs of the parameters, the values of K-S statistic,
\textit{p}-value, -2log(L), AIC, AD and CM of the second data set
are listed in Table 2. From these values, we note that the EWG model
is better than the WG, EW, GE and Weibull distributions in terms of
fitting to this data.

\begin{table}[htp!]
\centering \caption{MLEs(stds.), K-S statistics, \textit{p}-values,
$-2\log(L)$ and AIC for the strengths of 1.5 cm glass fibres.}
\begin{small}
\begin{tabular}{|l|lcccccc|}
\hline
Dist.& MLEs(stds.) & K-S  & \textit{p}-value &$-2\log(L)$& AIC& AD& CM  \\
\hline EWL & $\begin{array}{l}
 \hat{\alpha}=2207.22 (1.369),
\hat{\beta}=3.889(8.392)\\
\hat{\gamma}= 0.916 ( 5.144), \hat{\theta}= 0.968 (9.41)
\end{array}$
& 0.075 &   0.8657&111.738&119.738&0.2545&0.1266 \\
EW & $\begin{array}{l}
 \hat{\alpha}=37.23( 79.419),
\hat{\beta}= 0.87( 0.757)\\
\hat{\gamma}=1.453(0.758)
\end{array}$
&0.0813&0.7993& 112.621 &118.621& 0.3369&0.1464\\
%\hline
GE &~ $\hat{\alpha}$= 218.232(111.014), $\hat{\beta}$= 1.945( 0.192)& 0.0888   & .703 &113.032&117.032&0.3553&0.150 3\\
%\hline
%\hline
GEL &~$\begin{array}{l}
 \hat{\alpha}=   1006.437( 1.763),
\hat{\beta}=2.888(9.553)\\
 \hat{\theta}=0.965(9.417)
\end{array}$
 &0.0789 &   0.8277  &    111.817  &    117.817& 6 0.269   & 0.1292  \\
%\hline
WL &~ $\begin{array}{l} \hat{\beta}= 0.301 (0.019),
\hat{\gamma}=5.043( 0.728)\\
\hat{\theta}= 0.01(  0.968)
\end{array}$

 & 0.0877 &    0.718 &    123.931  & 129.931& 5 0.933&  0.20623 \\
%\hline

Weibull &~  $\hat{\beta}$=0.301(0.007), $\hat{\gamma}$=5.049(0.455)&0.0876   &0.7192&23.914 &127.914&0.9326&0.20614\\
\hline
\end{tabular}
\end{small}
\end{table}

Using the likelihood ratio (LR) test, we test the null hypothesis
H0: WG versus the alternative hypothesis H1: EWG, or equivalently,
H0: $\alpha=1$ versus H1: $\alpha\neq 1$. The value of the LR test
statistic and the corresponding \textit{p}-value are 16.713 and
4.34e-05, respectively. Therefore, the null hypothesis (WG model) is
rejected in favor of the alternative hypothesis (EWG model) for a
significance level $>$ 4.34e-05. For test the null hypothesis H0: GE
versus the alternative hypothesis H1: EWG, or equivalently, H0:
$(\gamma,\theta)=(1,0)$ versus H1: $(\gamma,\theta)\neq(1,0)$, the
value of the LR test statistic is 14.149 (\textit{p}-value =
0.0008), which includes that the null hypothesis (GE model) is
rejected in favor of the alternative hypothesis (EWG model) for a
significance level $>$ 0.0008. For test the null hypothesis H0:
Weibull versus the alternative hypothesis H1: EWG, or equivalently,
H0: $(\alpha,\theta)=(1,0)$ versus H1: $(\alpha,\theta)\neq(1,0)$.
The value of the LR test statistic is 15.247 (\textit{p}-value =
0.0005), which includes that the null hypothesis (Weibull model) is
rejected in favor of the alternative hypothesis (EWG model) for any
significance level. We also test the null hypothesis H0: EW versus
the alternative hypothesis H1: EWG, or equivalently, H0: $\theta=0$
versus H1: $\theta\neq 0$, the value of the LR test statistic is
1.713 (\textit{p}-value = 0.1905), which includes that the null
hypothesis (EW model) is rejected in favor of the alternative
hypothesis (EWG model) for a significance level $>$ 0.1905, and for
any significance level $<$ 0.1905, the null hypothesis is not
rejected but the values of AD and CM in Table 2 show that the EWG
distribution gives the better fit to the second data set than EW
distribution.\\
 Plots of the estimated cdf and pdf function of the
EWG,WG, EW, GE and Weibull models fitted to these two data sets
corresponding to Tables 1 and 2 are given in Fig. 4. These plots
suggest that the EWG distribution is superior to the WG, EW, GE and
Weibull distributions in fitting these two data sets.

\section{Conclusion}
We propose a new four-parameter distribution, referred to as the EWG
distribution which contains as special sub-models the generalized
exponential-geometric (GEG), complementary Weibull-geometric (CWG),
complementary exponential-geometric (CEG), exponentiated
Rayleigh-geometric (ERG) and Rayleigh-geometric (RG) distributions.
The hazard function of the EWG distribution can be decreasing,
increasing, bathtub-shaped and unimodal. Several properties of the
EWG distribution such as quantiles and moments, maximum likelihood
estimation procedure via an EM-algorithm, R\'{e}nyi and Shannon
entropies, moments of order statistics, residual life function and
probability weighted moments are studied. Finally, we fitted EWG
model to two real data sets to show the potential of the new
proposed distribution.

\end{document}